\documentclass[
  journal=largetwo,
  manuscript=article-type,
  year=2025,
  volume=37,
]{cup-journal}
\usepackage{amsmath, amssymb}
\usepackage[nopatch]{microtype}
\usepackage{booktabs}
\usepackage{subcaption}
\usepackage{multirow}

\usepackage{acronym}
\acrodef{DIP}[\texttt{DIP}]{\texttt{Deep Imaging Pipeline}}
\acrodef{RMS}[RMS]{root mean squared error}
\acrodef{RFI}[RFI]{radio-frequency interference}
\acrodef{ASVO}[ASVO]{All-Sky Virtual Observatory}
\acrodef{SNR}[SNR]{\rm signal-to-noise~ratio }
\acrodef{GGSM}[GGSM]{GLEAM Global Sky Model}
\acrodef{PSF}[PSF]{point spread function}
\acrodef{GAMA}[GAMA]{Galaxy and Mass Assembly}
\acrodef{MIDAS}[MIDAS]{MWA Interestingly Deep AStrophysical}
\acrodef{GOLD}[GOLD]{GAMA23 OverwheLming Deep}
\acrodef{AGN}[AGN]{active galactic nuclei}
\acrodef{RLF}[RLF]{radio luminosity function}
\acrodef{WISE}[WISE]{Wide-field Infrared Survey Explorer}
\acrodef{LoTSS}[LoTSS]{LOFAR Two-metre Sky Survey}
\acrodef{SED}[SED]{spectral energy distribution}
\acrodef{FWHM}[FWHM]{full width at half maximum}

\title{The MWA Interestingly Deep AStrophysical (MIDAS) Survey: Survey Description and First Data Release}

\author{S. Paterson}
\affiliation{International Centre for Radio Astronomy Research, Curtin University, Bentley, 6102, WA, Australia}
\email[S. Paterson]{sean.paterson@icrar.org}

\author{N. Seymour}
\affiliation{International Centre for Radio Astronomy Research, Curtin University, Bentley, 6102, WA, Australia}

\author{N. Hurley-Walker}
\affiliation{International Centre for Radio Astronomy Research, Curtin University, Bentley, 6102, WA, Australia}

\author{T. J. Galvin}
\affiliation{CSIRO Space \& Astronomy, Commonwealth Scientific and Industrial Research Organisation, Bentley, 6102, WA, Australia}

\author{S. W. Duchesne}
\affiliation{CSIRO Space \& Astronomy, Commonwealth Scientific and Industrial Research Organisation, Bentley, 6102, WA, Australia}

\author{J. Morgan}
\affiliation{CSIRO Space \& Astronomy, Commonwealth Scientific and Industrial Research Organisation, Bentley, 6102, WA, Australia}

\author{T. M. O. Franzen}
\affiliation{SKA Observatory, Jodrell Bank, Lower Withington, Macclesfield SK11 9FT, UK}


\begin{document}

\begin{abstract}
We present the deepest MWA survey to date, targeting the GAMA9 and GAMA23 survey fields at 215.68\,MHz. With a bandwidth of 30.72\,MHz, the two images cover ${\sim}1250\text{\,deg}^2$ for the GAMA9 field and ${\sim}850\text{\,deg}^2$ for the GAMA23 field, with an angular resolution of ${\sim}60''$. The GAMA9 image was formed from 1~282 2-min snapshot observations (42.7\,hours) and the GAMA23 image from 1~701 observations (56.7\,hours) achieving depths of $0.923 \pm 0.003$ and $0.331 \pm 0.001$\,mJy/beam at the field centres respectively. We processed the data with an updated version of the GLEAM-X pipeline tuned for deep surveys and make this MWA {\tt Deep Imaging Pipeline} code public. Updates to the pipeline include automation of the observation downloading, calibration, and imaging along with optimisation of the \texttt{CLEAN}ing methods and parameters. Above $5\sigma$, we catalogue 53~419 and 80~407 radio components in the GAMA9 and GAMA23 fields respectively. We find that the completeness corrected source counts for both fields are in good agreement down to $25$--$50$\,mJy with those from LOFAR and earlier MWA surveys. We demonstrate the utility of this survey by highlighting a few sources of particular interest and describing future science possible with this survey.
\end{abstract}

\section{Introduction}

The low-frequency radio sky has become a new frontier over the last decade with numerous observations from facilities such as the LOw-Frequency ARray (LOFAR; \citealt{vanhaarlem2013}), the Murchison Widefield Array (MWA; \citealt{tingay2013}) and the upgraded Giant Metrewave Radio Telescope (uGMRT; \citealt{gupta2017}). At lower radio frequencies, emissions evolve on relatively long time scales \autocite{franzen2021}, assisting in the study of remnant radio galaxies \autocite{quici2021}, galaxy cluster relics and halos \autocite{george2015, george2017}, supernova remnants \autocite{mantovanini2024}, and many other phenomena. Emissions at these frequencies originate from the lobes of \ac{AGN}, rather than the core; this negates beaming effects and local interactions \autocite{franzen2021}, enabling reliable measurements for average jet powers. In addition, well-calibrated low-frequency radio data offers extinction-free methods to evaluate star formation rates \autocite{beswick2015, heesen2019}. 

The study of the universe at lower radio frequencies continues to advance with the Square Kilometre Array Observatory's (SKAO)\footnote{https://www.skao.int/en} Low telescope under construction, and thanks to pathfinders such as LOFAR and precursors such as the MWA currently in operation, a greater number of low-frequency radio surveys are being conducted. The LOFAR Two-metre Sky Survey (LoTSS; \citealt{shimwell2017, shimwell2019, shimwell2022}) is an ongoing survey of the northern sky in the 120--168\,MHz range. At the time of writing, the second data release covers 27\% of the northern sky over two regions with areas of $4~178\text{\,deg}^2$ and $1~457\text{\,deg}^2$ respectively \autocite{shimwell2022}. The resulting Stokes \textit{I} image achieves a median \ac{RMS} of 83\,$\mu$Jy/beam with a resolution of 6$''$. From this they have produced a catalogue containing 4~396~228 radio sources.

For the southern sky, the GaLactic and Extragalactic All-sky MWA (GLEAM; \citealt{wayth2015}) survey produced a catalogue of 307~455 radio sources over a $24~831\text{\,deg}^2$ area in the frequency range 72--231\,MHz \autocite{hurley-walker2017} using observations conducted with the MWA in its Phase~\textsc{i} configuration. This catalogue was extended with a second GLEAM data release producing a compact source catalogue of 22~037 components for an area that covered ${\sim}50$\% of the Galactic plane \autocite{hurley-walker2019}. The MWA was then upgraded moving it to ``Phase~\textsc{ii}'' with the ``extended'' configuration which offered increased sensitivity and a higher angular resolution with longer baselines and improvements in their distribution \autocite{wayth2018}, With this configuration the GaLactic and Extragalactic All-sky MWA eXtended (GLEAM-X; \citealt{hurley-walker2022}) survey observed the sky south of Dec $+30^\circ$ across a frequency range of 72 to 231\,MHz; the initial data release over an area of $1~447\text{\,deg}^2$ produced a catalogue of 78~967 sources. The second GLEAM-X data release provides a catalogue of an additional 624~866 radio sources from a processed area of $12~892\text{\,deg}^2$ \autocite{ross2024}. Most recently, a third GLEAM-X data release utlised a joint deconvultion technique to combine GLEAM and GLEAM-X data for an area covering a ${\sim}3~800\,\text{\,deg}^2$ of the southern Galactic Plane, delivering improved fidelity with a resulting source catalogue containing 98~207 radio sources \autocite{mantovanini2025}.

The \ac{GAMA} project was designed to study galaxy structures for low to intermediate redshift galaxies and evaluate the cold dark matter (CDM) model \autocite{driver2009}. It has two components, a spectroscopic survey and a compilation of imaging with photometry across the electromagnetic spectrum. \textcite{liske2015} reported an observational campaign using the Anglo-Australian Telescope for a ${\sim}286\,\text{\,deg}^2$ area across five regions collecting spectra with the AAOmega spectrograph. These spectra cover the wavelength range 3740–-8850\,Å with a pixel size of 1.04\,Å. The spectroscopic survey was expanded further by \textcite{driver2022} using publicly available data derived from \textit{GALEX}, ESO KiDS, ESO VIKING, \textit{WISE}, and \textit{Herschel} Space Observatory imaging for a slightly reduced area of ${\sim}250\,\text{\,deg}^2$ across the five regions. The resultant survey offers a wealth of data including, but not limited to, galaxy spectra, spectroscopic redshifts and stellar masses. It covers frequency ranges from X-ray (1\,nm) to microwave (1\,m) and at the time of writing has resulted in spectra for 248~682 galaxies and 330~542 redshifts with a redshift completeness of 95\% to $r_\mathrm{KiDS} = 19.65$\,mag.

Recently, the Evolutionary Map of the Universe (EMU; \citealt{norris2021}, \citealt{hopkins2025}) project has extended the data available for the 60\,deg$^2$ GAMA23 region into the radio band with observations at 888\,MHz, 936\,MHz and 1320\,MHz \autocite{leahy2019, gurkan2022} produced with the Australian SKA Pathfinder (ASKAP; \citealt{johnston2007}, \citealt{deboer2009}, \citealt{hotan2021}) telescope, a pathfinder to the SKAO's Mid telescope. EMU will ultimately cover the entire Southern sky at dec.$<0\,$deg including the other four GAMA fields. In addition, a wealth of derived products have been determined from the surveys including, but not limited to, stellar mass \autocite{driver2022} and AGN bolometric luminosities \autocite{thorne2021}. The introduction of low-frequency radio data with a greater sensitivity to these fields that are already well studied at higher frequencies opens up further opportunities to better understand the physical processes occurring within them.

The \ac{MIDAS} survey offers complementary data to GLEAM-X for selected GAMA regions. GLEAM-X provides low-frequency radio data for a wide-field view of the southern hemisphere, here we offer additional depth at 215.68\,MHz to the GAMA9 and GAMA23 regions through a concentrated observation strategy with ${\sim}70$--$90$ hours of integration per field.

We provide an outline of the MWA observations conducted for the \ac{MIDAS} survey that were processed for this data release in Section 2. In Section 3, we describe the pipeline and automation used to process the observational data. The observation selection and mosaicking process is described in Section 4. Then in Section 5 we outline the properties of the final images and the resulting data products along with a preliminary investigation into a few example sources exhibiting atypical radio emission to demonstrate the utility of the survey. Lastly, in Section 6 we summarise the results.

\section{Observational Data}

To add to the wealth of information available from the \ac{GAMA} project, we used the MWA radio telescope to conduct thousands of observations of the five \ac{GAMA} fields along with the Spitzer South Pole Telescope Deep Field. The observations were carried out over the same frequency range as GLEAM to offer complementary data to the GLEAM(-X) surveys \autocite{beardsley2019}. They were conducted in 2-min snapshots due to the fixed grid of points at which the MWA can form beams, i.e. it cannot track sources. This strategy also adds the benefit of being able to apply a non-varying beam pattern and ionospheric correction to each snapshot rather than multiple corrections that would be required for longer-duration observations. We used MWA's ``Phase~\textsc{ii} extended'' configuration as it maximises the angular resolution while minimising classical and sidelobe confusion. For a subset of day time observations, the survey made use of a Sun nulling technique that places the Sun in a low-sensitivity area between the sidelobes.

We focused our deep field imaging on two of the most observed fields from the \ac{GAMA} survey, GAMA9 and GAMA23. From the surveys, we selected observations centred at 215.68\,MHz with a bandwidth of 30.72\,MHz outlined in Table~\ref{table_obs}. This frequency offers a generous field of view and the addition of 215.68\,MHz data for these fields will complement the existing data, enabling more science opportunities.

Under the MWA project ID \texttt{G0047} during the period March 2018 to August 2018, 2136 observations (71.2 hours) were conducted of the \ac{GAMA}23 field. Despite the wide field of view of the MWA ($>600\,$deg$^2$), the main focus of these observations was the $60 \text{\,deg}^2$ of multi-wavelength and spectroscopic data between RA $+339^{\circ}$ to $+351^{\circ}$ and between Dec $-35^{\circ}$ to $-30^{\circ}$. This field is close to zenith for the MWA with very few bright sources in the field of view, providing ideal conditions for observing.

The second selected field, \ac{GAMA}9, is off-zenith with numerous bright sources in the field of view, including Hydra-A. This presented significant challenges for imaging discussed in Section~\ref{imaging}. These observations focused on the $60 \text{\,deg}^2$ area from RA $+129^{\circ}$ to $+141^{\circ}$ and Dec $-2^{\circ}$ to $+3^{\circ}$. The survey observed this field between April 2020 and September 2020 under MWA project ID \texttt{G0069}, producing 2657 observations (88.6 hours).

\begin{table}[t!]
\begin{threeparttable}
\caption{Description of the GAMA fields observed where the first row is the range of the RA for the observed GAMA field in degrees, the second row is the range of Dec for the field in degrees and the third is the area of region in square-degrees. The fourth row is the MWA project ID the observations were made under, the fifth is the dates the observational campaign started and the sixth is the date the campaign finished. The seventh row is the range of RA for the pointing centre of the observations in degree, the eighth row is Dec for the observation centres and the ninth is the approximate range for the areas observed (per observation). The tenth row is the average RMS in each snapshot measured by \texttt{BANE} after processing through the pipeline outlined in Sections \ref{section_dip}. The eleventh is the total number of observations conducted and the twelfth is the number of observations that were used for the final mosaics.}
\label{table_obs}
\begin{tabular}{lll}
\toprule
\headrow  & \multicolumn{2}{c}{Region}\\
\headrow  & \multicolumn{1}{c}{GAMA23} & \multicolumn{1}{c}{GAMA9}\\
\midrule
RA  range (deg) & $339$ to $351$ & $129$ to $141$ \\
\midrule
Dec  range (deg) & $-35$ to $-30$ & $-2$ to $3$\\
\midrule
Area of GAMA field (deg${^2}$) & $60$ & $60$ \\
\midrule
Project ID & \texttt{G0047} & \texttt{G0069} \\
\midrule
Observing start date & March 2018 & April 2020 \\
\midrule
Observing end date & August 2018 & September 2020 \\
\midrule
Observing RA centre (deg) & $335.7$ to $354.3$ & $124.1$ to $146.6$\\
\midrule
Observing Dec centre (deg) & $-40.3$ to $-26.8$ & $-11.1$ to $11.9$\\
\midrule
Observing area per pointing (deg${^2}$) & $\sim{}700$--$750$ & $\sim{}600$--$700$ \\
\midrule
Average snapshot RMS (mJy/beam) & 6 & 17 \\
\midrule
Total observations & $2~136$ & $2~657$ \\
\midrule
Observations mosaicked & $1~701$ & $1~282$ \\
\bottomrule
\end{tabular}
\end{threeparttable}
\end{table}

\section{Deep Imaging Pipeline}\label{section_dip}

To process the thousands of observations required for the deep field images, we created the \ac{DIP}\footnote{https://github.com/sjpaterson/dip}. \ac{DIP} is adapted from the pipeline used in the GLEAM-X survey\footnote{https://github.com/GLEAM-X/GLEAM-X-pipeline} and utilises a combination of Nextflow \autocite{ditommaso2017}, Python and Bash scripts to automatically download the data from the \ac{ASVO}\footnote{https://asvo.mwatelescope.org} and then process it. The pipeline is primarily designed to process MWA data on the Setonix supercomputer node at the Pawsey Supercomputing Centre\footnote{https://pawsey.org.au}, however it can easily be ported to other high performance computing environments and is fully customisable allowing any selection of \texttt{CLEAN}ing\footnote{Imaging was completed with \textsc{WSClean} \autocite{offringa2014, offringa2017} detailed in Section~\ref{imaging}.} parameters with minor modification. The pipeline takes a list of MWA observation IDs and outputs a \texttt{CLEAN}ed image, RMS map, background map and source catalogue for each observation.

For this work, we used resources provided by the Pawsey Supercomputing Centre, including the Garrawarla and Setonix supercomputer nodes, processing ${\sim}120$\,observations per day and Acacia, a distributed disk-array, for storage. The processing is broken down into three main steps; calibration, imaging, and post-imaging. We describe these steps below, highlighting the modifications to the GLEAM-X pipeline.

While all the individual two-minute snapshot observations used here pointed at the GAMA fields in question, they all had different off-zenith electrically steered beams and different sky pointing centres. Hence, the data could not be combined in the $(u,v)$-plane or imaged in one go as the joint-deconvolution algorithms \autocite{vandertol2018} do not scale easily past ${\sim}20$\,observations \autocite{mantovanini2025}. Hence, each snapshot observation was imaged and \textsc{CLEAN}ed separately.

\subsection{Calibration}\label{Calibration}

Calibration was performed using the same methods as GLEAM-X. We utilised the \ac{GGSM} from the GLEAM-X pipeline that was derived primarily from GLEAM along with additional models for the brighter and more complex sources from literature \autocite{hurley-walker2022, ross2024}. This model provides the flux densities for the observed sources from which we derive our per-frequency channel, per-polarisation, per-antenna bandpass calibration solutions.

For each observation, the model was cropped to the 250 brightest sources after applying an estimate of the primary beam attenuation predicted by the Full Embedded Element (FEE; \citealt{sokolowski2017}) response within a radius of $30^{\circ}$ from the pointing centre. \texttt{MitchCal} \autocite{offringa2016} was then used to generate the calibration solution for each measurement set.

A phase slope was generated for the solution by dividing through by a selected reference antenna, the last tile in the array, number 128. The phase slope per polarisation for each MWA station calibration solution was inspected\footnote{Using a modified version of https://github.com/MWATelescope/mwa-calplots to automatically detect potentially problematic tiles available at https://github.com/sjpaterson/mwa-calplots.}. A linear trend was fitted; if the RMS of the residual was $>$20\%, the entire station was automatically flagged for the remainder of the observing night. If one or more tiles were flagged, a new calibration solution was generated. If more than 50~bad tiles were detected, the processing was terminated and the observation was flagged as unsuccessful. A total of 131 observations were flagged unsuccessful in GAMA23 and 179 observations in GAMA9.

Additionally, if more than 25\% of the solutions were flagged due to \ac{RFI}, failure to converge, or unmodelled sources in the sidelobes), the processing was halted and the observation was flagged as unsuccessful. This criteria did not flag any observations in GAMA23, but flagged an additional 22 observations in GAMA9. Once all quality checks were passed, the calibration solution was applied to the measurement set.

During the ASVO download, potential \ac{RFI} is flagged; for the final calibration step we rerun the \ac{RFI} flagging by calculating the running mean and standard deviation for the $(u,v)$-distance vs amplitude for the visibilities. If there were data points that were $3\sigma$ above the computed running mean, the baselines were flagged completely.

\subsection{Imaging}\label{imaging}

The imaging was performed with \textsc{WSClean} utilising a similar method to the GLEAM-X pipeline, except refining arguments to reduce the \ac{RMS} noise and improve deep imaging results. For the \texttt{CLEAN}ing, we moved from the $w$-stacking method used in GLEAM-X to $w$-gridding. Rather than assigning visibilities to the nearest $w$-plane, this method grids them to a small range of weighted $w$-planes \autocite{offringa2014, arras2021, ye2022}, significantly improving the data reduction speed with no penalty to image fidelity. We then followed the GLEAM-X method and imaged four separate sub-bands, each with a bandwidth of 7.68\,MHz and \texttt{CLEAN}ed the full integrated 30.72\,MHz wide band image. However, as our focus was on the final wideband image, we did not implement spectral fitting on the sub-bands that would enforce smooth spectra.

As observations move off-zenith, the shape of the primary beams differ for each polarisation. The further a pointing is from zenith, the greater the difference in response from the different dipole orientations, and therefore the greater the intrinsic polarisation of the primary beam. With the varying elevation of the fields, different pointings were required which resulted in different primary beam shapes. Therefore, we followed a similar method as the original GLEAM survey \autocite{hurley-walker2017} and imaged the XX and YY linear polarisations separately. We used the \texttt{link-polarization} option to ensure that any component detected in either polarisation was \texttt{CLEAN}ed in both and applied the primary beam to form a pseudo-Stokes \textit{I} image in the post-imaging stage (Section~\ref{post-image}).

\subsubsection{Noise Reduction}

In the final deep image, there are expected to be three main sources of noise: thermal, sidelobe and confusion. The thermal noise for the MWA is dominated by the sky noise with a small contribution from the receiver, while the sidelobe noise is additional noise introduced from the synthesised beam's sidelobes of residual un\texttt{CLEAN}ed sources, both faint sources in the primary beam and potentially brighter sources in the grating lobes of the telescope \autocite{franzen2016}. The confusion noise is the fluctuations in the image caused by the collective contribution of all sources within the main lobe of the synthesised beam where, as the number of observations stacked increases, the number of sources observed increases until the classical confusion limit is reached and the density of the sources become so great that they cannot be resolved \autocite{condon1974, franzen2016}.

As confusion and sidelobe noise were expected to dominate the final mosaicked images, we selected a uniform weighting scheme. Uniform weighting provides a smaller beam minimising confusion noise and a more Gaussian main lobe response minimising sidelobe noise. These advantages are partly offset by a higher thermal noise due to the down-weighting of short baselines.

To further reduce sidelobe noise, a cosine-weighted window (Tukey) function \autocite{harris1978} was applied to the inner $(u,v)$-plane to smooth out the sensitivity at short baselines to minimise `pedestals' in the dirty beam. We found that an inner Tukey window of $0\lambda$ to $875\lambda$ offered a good trade-off between sensitivity and rejection of large-scale structure.

\subsubsection{Robust Clean Masking}\label{Masking}

A full run of \ac{DIP} was initially run on the GAMA9 and GAMA23 fields using an \texttt{auto-mask} threshold of $3\sigma$. The resulting outputs were the \texttt{CLEAN}ed images, RMS maps produced by \texttt{BANE} \autocite{hancock2012, hancock2018} and source catalogues produced by \texttt{AEGEAN}\footnote{Throughout this article, source finding was performed using the default settings for \texttt{AEGEAN} which can be obtained from https://github.com/PaulHancock/Aegean.} \autocite{hancock2012, hancock2018} for each snapshot. Using this level of \texttt{auto-mask} cleans the image well, but as a side effect it deconvolves noise peaks removing flux from real sources as described by \textcite[Appendix~A.2.]{duchesne2025a}\footnote{Similar tests were performed for MIDAS data and showed similar results, with further tests verifying that the FITS masking approach reduced the issue.}. These errors accumulate when stacking snapshots resulting in a large flux density offset for all detected sources when compared to the reference catalogue. To prevent this from occurring, \ac{DIP} was rerun using a \texttt{CLEAN}ing mask (\texttt{fits-mask}) to attempt to clean only real sources.

To generate the mask, we created mosaics from the the $3\sigma$ auto-mask \texttt{CLEAN}ed snapshots and then performed blind source finding on them with \texttt{AEGEAN}. While the flux densities in the source catalogues are incorrect, the sources detected and their positions are real. 

We utilised the auto-masked \texttt{CLEAN}ed snapshots and RMS maps, selecting all pixels above $3\sigma$ and creating island maps. We then referenced the source positions in the catalogue produced from the mosaic and removed any island that did not have a corresponding source associated to it. The final reduced island maps were then used for the \texttt{fits-mask} parameter during the re-processing of the snapshots.

\subsubsection{Other Imaging Parameters}

Through the use of the \texttt{wgridder} option in \textsc{WSClean}, we found an approximate ${2.5}\times$ speed increase over the $w$-stacking method. This speed improvement made available additional CPU cycles to fine-tune the imaging options and further improve the \texttt{CLEAN}ing process. We investigated altering the number of sub-bands imaged, the auto-threshold limit, the multi-scale gain, the maximum number of major \texttt{CLEAN}ing iterations (\texttt{nmiter}), the peak flux density reduction per iteration (major iteration gain referred to as \texttt{mgain}) and the image size.

To gauge the behaviour of the selected parameters for our visibilities, we \texttt{CLEAN}ed a small number of observations from GAMA9 using a number of different parameter combinations. A collection of ten observations that represented different imaging conditions were used to provide a consistent comparison through this parameter fine-tuning process. Altering the multi-scale gain from 0.05 to 0.25 did not result in a significant change in the image \ac{RMS}. Variations of \texttt{mgain} between 0.6 and 0.95 along with \texttt{nmiter} between 5 and 20 revealed an optimal reduction of the image \ac{RMS} for a minimal increase in computational time with an \texttt{mgain} value of 0.75 and a \texttt{nmiter} value of 10. The \texttt{mgain} change resulted in a decrease of RMS by 2 to 5\% percent without any significant change in compute time (less than a minute), while the \texttt{nmiter} change saw a 10\% improvement in RMS with an extra ${\sim}60$ minute compute time. Increasing the number of sub-bands from 4 to 8 and 16 resulted in a reduction in the image \ac{RMS} by 2 to 6\% but required a 60 to 150 minute increase in processing time. Increasing the image size while maintaining the pixel scale \texttt{CLEAN}ed a larger area that resulted in an image \ac{RMS} reduction with a small increase in processing time. Although some parameters increase storage requirements, we had sufficient resources to meet those requirements. Therefore, for our final parameter values we chose to decrease the \texttt{mgain} from 0.85 used in the GLEAM-X pipeline to 0.75, increased \texttt{nmiter} from 5 to 10 and increased the image size from 8~000\,pixels $\times$ 8~000\,pixels to 10~000\,pixels $\times$ 10~000\,pixels.

As a result, the \textsc{WSClean} parameters listed in Table~\ref{table_wsclean_parameters} including a comparison with GLEAM-X, were selected for the image reduction process.

\begin{table}[t!]
\begin{threeparttable}
\caption{Comparison of the parameters used for \textsc{WSClean} between GLEAM-X and DIP. The first column is the parameter, the second column is the values used in GLEAM-X and third column is the values used in DIP. The `scale' parameter is dependant on the central frequency of the observations; here we provide the value used for the processing on channel 169 (215.68\,MHz). These are the parameters used for imaging the snapshots that are mosaicked for the final deep field images with the fits-mask input described in detail in Section \ref{Masking}.}
\label{table_wsclean_parameters}
\begin{tabular}{lll}
\toprule
\headrow Parameter & GLEAM-X & DIP\\
\midrule
\texttt{wgridder} & No & Yes \\ 
\midrule
\texttt{multiscale} & Yes & Yes \\
\midrule
\texttt{mgain} & 0.85 & 0.75 \\
\midrule
\texttt{multiscale-gain} & 0.15 & 0.15 \\
\midrule
\texttt{nmiter} & 5 & 10 \\
\midrule
\texttt{niter} & 10000000 & 10000000  \\
\midrule
\texttt{auto-mask} & 3 & Not Used \\
\midrule
\texttt{fits-mask} & Not Used & Yes \\
\midrule
\texttt{auto-threshold} & 1 & 1 \\
\midrule
\texttt{size} & 8000 8000 & 10000 10000 \\
\midrule
\texttt{scale} & 0.0035503 & 0.0035503 \\
\midrule
\texttt{weight} & briggs 0.5 & uniform \\
\midrule
\texttt{taper-inner-tukey} & Not Used & 875 \\
\midrule
\texttt{minuv-l} & Not Used & 0 \\
\midrule
\texttt{pol} & I & XX,YY \\
\midrule
\texttt{link-polarizations} & No & Yes \\
\midrule
\texttt{fit-spectral-pol} & 2 & Not Used \\
\midrule
\texttt{reuse-primary-beam} & Yes & No \\
\midrule
\texttt{apply-primary-beam} & Yes & No \\
\midrule
\texttt{join-channels} & Yes & Yes \\
\midrule
\texttt{channels-out} & 4 & 4 \\
\bottomrule
\end{tabular}
\end{threeparttable}
\end{table}

\subsection{Post Imaging}\label{post-image}

\texttt{CLEAN}ing resulted in images of the XX and YY polarisations with angular resolutions of ${\sim}1'$. The Stokes \textit{I} ($I_I$) image was formed using the methods implemented by \textcite{sault1996} and \textcite{morgan2022} with

\begin{equation}
\label{Stokes_I_Equation}
I_I(l) = \frac{\sum_p A(p) B(l,p) I(l,p) / \sigma^2(p)}{\sum_p A^2(p) B^2(l,p) / \sigma^2(p)},
\end{equation}

where $p$ is the linear polarisation (XX and YY), $A(p)$ is a direction-independent scaling correction, $B(l,p)$ is the sky direction and polarisation-dependent primary beam power pattern, $I(l,p)$ is the intensity of each polarisation and $\sigma$ is the \ac{RMS} at the centre of the beam for the primary beam corrected polarisation image, which like the scaling correction is assumed to be direction-independent.

The primary beam power for the XX and YY polarisations were calculated using the \texttt{mwa\_pb\_lookup} package\footnote{https://github.com/johnsmorgan/mwa\_pb\_lookup} \autocite{morgan2021} which utilises the beam model created by \textcite{sokolowski2017}. This software generates a beam intensity image based on the imaged polarisations providing a one-to-one pixel map for the beam power ($B(l,p)$).

The beam intensity images were applied to each linear polarisation to correct for the primary beam and \texttt{BANE} was run to estimate the \ac{RMS}. The beam intensities were used to determine the location of peak sensitivity and the \ac{RMS} value ($\sigma$) at this location for each polarisation was recorded. To determine the scaling factor $A(p)$, we performed a blind source search on each primary beam corrected polarisation image using the source finding software \texttt{AEGEAN}. To ensure that the most reliable sources were selected, the source catalogues were reduced to isolated sources that did not have a neighbour within a $6'$ cross-matching radius, had a flux density $>0.5$\,Jy, and were matched to a source in the \ac{GGSM} sparse unresolved catalogue with a maximum separation distance to the catalogue source of $1'$. The integrated flux density ratios between the filtered sources in each polarisation and the corresponding sources in the \ac{GGSM} catalogue were calculated. The weighted average integrated flux density ratio was calculated for each linear polarisation using the \ac{SNR} of the MIDAS sources, down-weighting sources with lower SNRs. This resulted in an average scaling factor ($A(p)$) of $0.91 \pm 0.11$ for GAMA9 and $1.07 \pm 0.02$ for GAMA23.

The Stokes \textit{I} image was formed using equation \ref{Stokes_I_Equation}, the \ac{RMS} and background was estimated with \texttt{BANE} and a source catalogue was produced by \texttt{AEGEAN}. A quality check for a minimum of 500 sources was applied; any observation that did not pass this check was rejected. To correct source position offsets introduced in the snapshot images due to ionospheric distortions, we utilised \texttt{FITS\_WARP}\footnote{https://github.com/nhurleywalker/fits\_warp} which calculates the source position offsets between the observational source catalogue and a reference catalogue with accurate source locations, fits a general non-parametric model to the vector offsets using a radial basis function and then applies a de-distortion correction to the observation in the image plane \autocite{hurley-walker2018}. Before \texttt{FITS\_WARP} received the observation source catalogue, it was reduced to isolated sources that did not have a neighbouring source within 6$'$ to avoid source confusion during the matching process. For the reference catalogue, we utilised the positional-based catalogue of unresolved sources generated from the NRAO VLA Sky Survey (NVSS; \citealt{condon1998}) and the {Sydney University Molonglo Sky Survey (SUMSS; \citealt{bock1999, mauch2003, murphy2007}) described in detail by \textcite{hurley-walker2022}. 

Previous flux scaling corrections apply a single multiplicative factor to the entire image. To determine and correct systematic spatial flux density scale errors we utilise \texttt{flux\_warp}\footnote{https://gitlab.com/Sunmish/flux\_warp} \autocite{duchesne2020}, parsing a new source catalogue generated with \texttt{AEGEAN} along with the \ac{GGSM} sparse unresolved catalogue. This detects the differences in source flux density between the catalogues and corrects the image using a user selectable mode. We selected the quadratic screen mode because it provided a good fit to our data while providing a robustness to outliers. This screen fits a two-dimensional second order polynomial model to the differences in the flux density scale differences and applies a correction. The resulting corrected output image was the final image considered for use in the deep field mosaic. From this final image, the final \ac{RMS} and catalogue were produced with \texttt{BANE} and \texttt{AEGEAN}, respectively.

\subsection{Final Quality Control}

In the initial processing run, if an observation fails, often due to random compute/disk errors, it is reattempted for a maximum of three attempts without change. However, after the initial processing run, a second run was initiated for all observations that failed the initial three attempts. The second run used a different reference tile for the calibration outlined in Section~\ref{Calibration}. The selected reference tile was changed from the last tile in the array, number 128, to the first tile in the array, number 1.

After switching the reference tile, 1~025 of the 2~657~GAMA9~observations were reprocessed in which 392 failed. For GAMA23, 322 of the 2~136 observations were reprocessed with 210 failing. Each successfully processed observation was then visually inspected in which 221~GAMA9 observations showed significant imaging artefacts and were excluded from consideration for the final mosaic. The distortion issues were not investigated further as thousands of observations were successfully processed and it was unlikely that recalibrating this small group of observations would provide a significant improvement to the final mosaic. No problematic observations were observed for GAMA23. This resulted in a total of 2044 valid observations for GAMA9 and 1926 observations for GAMA23, as presented in Table~\ref{table_processingresults}.

\begin{table}[t!]
\begin{threeparttable}
\caption{The number of observations that were successfully processed. The first column is the description of the processing stage, the second column is the results for GAMA9 and the third column is the results for GAMA23.}
\label{table_processingresults}
\begin{tabular}{lll}
\toprule
\headrow Processing Stage & GAMA9 & GAMA23\\
\midrule
Total Number of Observations & 2~657 & 2~136 \\
\midrule
Successfully Processed with First Reference Tile & 1~632 & 1~814 \\
\midrule
Successfully Processed with Last Reference Tile & 633 & 112 \\
\midrule
Failed Processing & 392 & 210 \\ 
\midrule
Failed Visual Inspection & 221 & 0 \\ 
\midrule
Total Successfully Processed & 2~044 & 1~926 \\ 
\bottomrule
\end{tabular}
\end{threeparttable}
\end{table}

\section{Mosaicking}\label{section_mosaic}
On completion of the processing of the observational data with \ac{DIP}, the resulting images were co-added to produce the deep field images for each GAMA field. To produce a more consistent and predictable \ac{PSF} and improve angular resolution, we select observations based on restoring beam dimensions. To keep the noise low there is a trade-off between smaller beam size or a larger number of images. With this large number of observations, where confusion noise is important, co-addition of all images can have a negative effect, potentially increasing the final noise level. Therefore, it was necessary to attempt to find the balance by trying to select only observations that would improve the noise level in the final deep image while optimising the beam size to improve the resolution in the final mosaics.

\subsection{Beam Cuts}\label{beam_cuts}

\begin{figure*}[t!]
\centering
\includegraphics[width=0.49\linewidth]{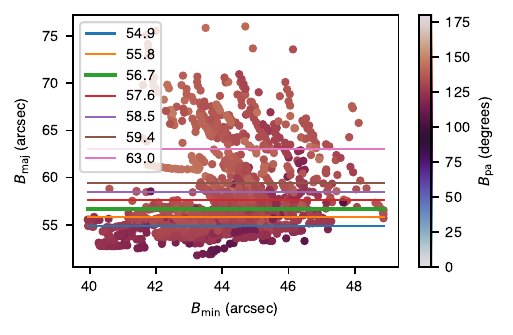}
\includegraphics[width=0.49\linewidth]{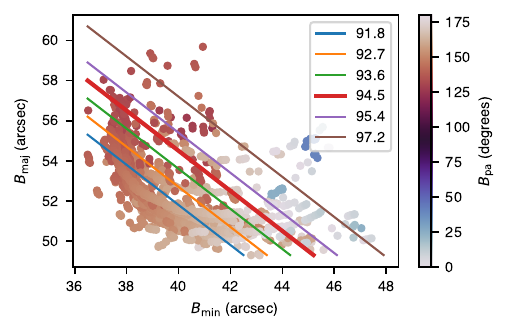}
\caption{Beam cuts (solid lines) applied on the GAMA9 observations (left) and GAMA23 (right) where the x-axis represents the minor axis for the beam and the y-axis represent the major axis. For GAMA9, the cuts were applied on the $B_\mathrm{maj}$, where the cut at 56.7$''$ (green line) produced the optimum final RMS. For GAMA23, the beam cuts were based on the linear equation $B_\mathrm{maj} = -B_\mathrm{min} + C$ where the cut at $C=94.5''$ (red line) produced the optimum final RMS.}
\label{fig_beaminfo}
\end{figure*}

To optimise the beam size for the final mosaic of each region, the observations were selected based on the dimensions of the Gaussian fits to the synthesised beam referred to as the restoring beam. A smaller beam size enables an improvement in angular resolution and in \ac{RMS} due to lower classical confusion. Hereafter, we refer to this filtering process based on restoring beam dimensions as `beam cuts'. Based on the distribution of the \ac{FWHM} beam parameters shown in Figure~\ref{fig_beaminfo}, we decided to cut larger beams in the following way. For GAMA9, as the beam is fairly elongated with a narrow range of restoring beam position angles ($B_\mathrm{pa}$), we decided to cut on the restoring beam's major axis ($B_\mathrm{maj}$) only as described in Table~\ref{table_gama9beamcuts}. For GAMA23 with a rounder beam and a wider range of position angles we decided to cut on maximum beam size, i.e. a diagonal in $B_\mathrm{maj}$ v $B_\mathrm{min}$ parameter space  ($B_\mathrm{maj} = -B_\mathrm{min} + C$) outlined in Table~\ref{table_gama23beamcuts}, where $B_\mathrm{min}$ is the minor axis for the restoring beam. Finally, the mosaic for each beam cut was produced in the same fashion as the final images (as outlined in Section~\ref{coaddition}) and the \ac{RMS} estimated by \texttt{BANE} is presented in Table~\ref{table_gama23beamcuts}\footnote{We originally attempted to convolve observations to a common beam, however this introduced an offset into the flux density scale and therefore was not performed in the final deep field mosaics.}. The average \ac{RMS} measured within each GAMA region was then used to evaluate the effectiveness of the beam cut.

\begin{table}[t!]
\begin{threeparttable}
\caption{GAMA9 beam cuts where the first column is the $B_\mathrm{maj}$ for the cut, the second is the number observations selected from the cut to the 2~044 available observations and the third is the average RMS measured within the GAMA9 region.}
\label{table_gama9beamcuts}
\begin{tabular}{lll}
\toprule
\headrow $B_\mathrm{maj}$ (arcsec) & \# Obs Selected & Mean RMS (mJy/beam) \\
\midrule
54.9 & 591 & 0.797 \\ 
\midrule
55.8 & 1~040 & 0.769 \\ 
\midrule
56.7 & 1~282 & 0.755 \\ 
\midrule
57.6 & 1~450 & 0.760 \\ 
\midrule
58.5 & 1~536 & 0.760 \\ 
\midrule
59.4 & 1~642 & 0.762 \\ 
\midrule
63.0 & 1~820 & 0.778 \\ 
\bottomrule
\end{tabular}
\end{threeparttable}
\end{table}

\begin{table}[t!]
\begin{threeparttable}
\caption{GAMA23 beam cuts where the first column is the value of the constant $C$ in the linear equation $B_\mathrm{maj} = -B_\mathrm{min} + C$ used to filter the observations. The second column is the number of observations selected from the cut to the 1~926 available observations and the third column is the average \ac{RMS} measured within the GAMA23 region.}
\label{table_gama23beamcuts}
\begin{tabular}{lll}
\toprule
\headrow $C$ & \# Obs Selected & Mean RMS (mJy/beam) \\
\midrule
91.8 & 865 & 0.349 \\ 
\midrule
92.7 & 1~295 & 0.335 \\ 
\midrule
93.6 & 1~610 & 0.333 \\ 
\midrule
94.5 & 1~701 & 0.331 \\ 
\midrule
95.4 & 1~792 & 0.334 \\ 
\midrule
97.2 & 1~867 & 0.360 \\ 
\bottomrule
\end{tabular}
\end{threeparttable}
\end{table}

For GAMA9, the optimal beam cut with the lowest RMS was determined to occur at $B_\mathrm{maj}$ of 56.7$''$. This resulted in a total of 1~282 observations being included in the mosaic, representing 62.7\% of the available observations and equates to 42.7 hours of integration time. For the GAMA23 region, the optimal beam cut resulting in the lowest RMS occurred for a linear cut of $B_\mathrm{maj} = -B_\mathrm{min} + 94.5$. This resulted in 1~701 observations (88.3\% of the total available) selected for the mosaic, equating to an integration time of 56.7 hours.

\subsection{Co-addition}\label{coaddition}

The observations were reprojected and regridded using \texttt{SWARP} \autocite{bertin2002} with a \texttt{ZEA} projection. Within the same operation the observations were co-added together utilising the RMS maps generated by \texttt{BANE} for each observation to calculate the pixel weighting and produce the deep field images. This co-addition step is near-equivalent to applying Equation~\ref{Stokes_I_Equation} to all linear polarisations; however, utilising \texttt{SWARP} automates the resampling and re-projection.

The resulting GAMA23 mosaic from the MIDAS survey covered an area ${\sim}850 \text{\,deg}^2$ centred on GAMA23. Due to the larger area of the individual snapshots and a larger scatter in pointing centres, the GAMA9 mosaic covered a slightly larger area of ${\sim}1~250 \text{\,deg}^2$ centred on GAMA9. The image sizes for the mosaics were 9~316\,pixels $\times$ 7~301\,pixels for GAMA23 and 9~980\,pixels $\times$ 9~980\,pixels for GAMA9.

\subsection{Calculating the Point-Spread-Function}\label{psf}

Due to the varying restoring beam shapes and sizes for the individual observations, the \ac{PSF}s for the MIDAS mosaics while being very close to Gaussian, have an unpredictable shape. Therefore, we estimated the \ac{PSF} using the same methods as described by \textcite{hurley-walker2017}, measuring the shapes and sizes of known unresolved sources to determine the effective PSF dimensions. To ensure the final PSFs could be approximated as Gaussian, we stacked 2D models of all restoring beams obtained by \textsc{WSClean} for each observation within a MIDAS mosaic. We fit a 2D Gaussian to the resulting averages for GAMA9 and GAMA23. We subtracted the fitted 2D Gaussians from the MIDAS average beams and inspected the residuals which had a peak $<1$\% of the beam. This demonstrated that the final expected beam for each mosaic can be well described by a 2D Gaussian.

For each mosaic, we generated a source catalogue with \texttt{AEGEAN} and matched to the higher-resolution NVSS and SUMSS catalogue of unresolved sources described in Section~\ref{post-image}, filtering for sources with an \ac{SNR} $> 10$. We then created the PSF maps by calculating a HEALPix \footnote{Hierarchical Equal Area isoLatitude Pixelation \autocite{gorski2005} available from http://healpix.sourceforge.net/} map of the average measured source dimensions. The resulting PSF maps were used for source finding with \texttt{AEGEAN} on the MIDAS mosaics outlined in Section~\ref{sourcefinding}.

\subsection{Ionospheric Corrections}\label{blurring}

Small residual ionospheric effects in each observation can slightly vary source positions and introduce blurring into the mosaics. Although the changes in the shape of the PSF resulting from the blurring are measured in Section \ref{psf}, it results in an underestimation of the measured peak flux density (while conserving total flux density). Following the methods used by \textcite{hurley-walker2022}, we correct this effect by normalising the flux density scales in the MIDAS mosaics. We created maps for the theoretical PSFs for each deep field image by mosaicking the PSF maps for the reprojected snapshots generated in Section~\ref{coaddition}. \texttt{AEGEAN} was run using the theoretical PSFs to measure source dimensions; the resulting catalogues were filtered to bright unresolved sources through the methods used in Section~\ref{psf}. A positional ``blurring'' factor was calculated from the catalogues by 

\begin{equation}
\label{PSF_Blur}
A = \frac{F_\mathrm{maj} F_\mathrm{min}}{T_\mathrm{maj} T_\mathrm{min}}
\end{equation}

where $F_\mathrm{maj}$ and $F_\mathrm{min}$ represent the \ac{FWHM} of the major and minor axes fitted to the sources, and $T_\mathrm{maj}$ and $T_\mathrm{min}$ represent the \ac{FWHM} of the major and minor axes of the theoretical PSF. This resulted in positional blurring maps for each MIDAS mosaic shown in Appendix \ref{app_blur}, where the average blurring factor for GAMA9 was 1.10 and GAMA23 was 1.15. The MIDAS mosaics were then multiplied by their respective blurring map, bringing the integrated and peak flux densities into alignment.

\section{Results}

The processing resulted in a ${\sim}1250 \text{\,deg}^2$ deep field mosaic centred around GAMA9 (RA $+135^{\circ}$, Dec $+0.5^{\circ}$) and a ${\sim}850 \text{\,deg}^2$ deep field mosaic centred around GAMA23 (RA $+345^{\circ}$, Dec $-32.5^{\circ}$). In Figure \ref{fig_gamacutouds} we present a cutout of the resulting 60\,deg$^2$ GAMA9 and GAMA23 regions from the MIDAS mosaics. 

\begin{figure*}[ht!]
\centering
\includegraphics[width=0.99\linewidth]{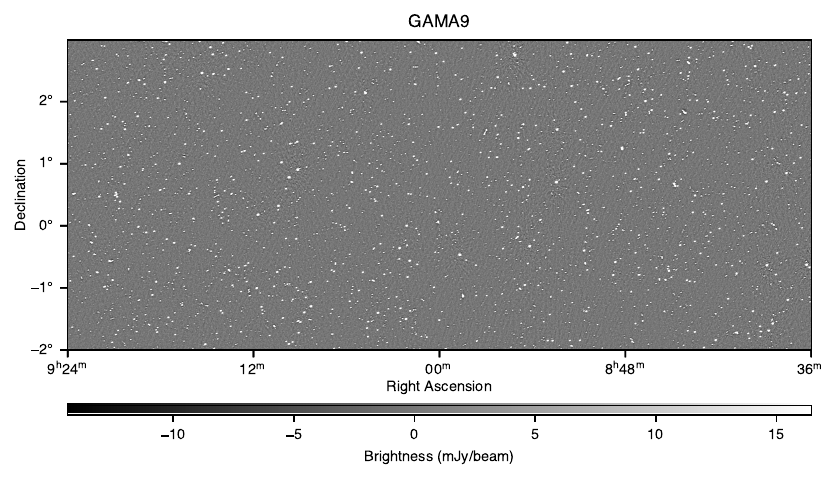}\\
\includegraphics[width=0.99\linewidth]{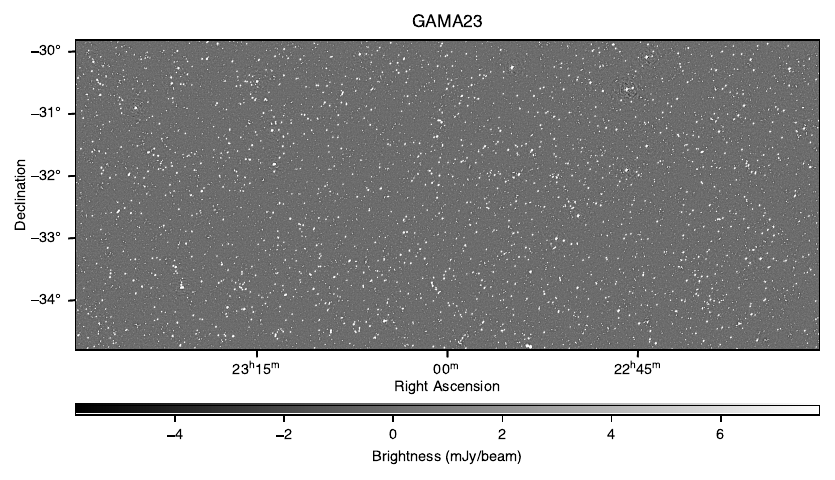}
\caption{Greyscale cutouts of the 60\,deg$^2$  GAMA9 region (top) and 60\,deg$^2$  GAMA23 region (bottom) from the MIDAS mosaics.}
\label{fig_gamacutouds}
\end{figure*}

\subsection{Noise Analysis}\label{rms}
\subsubsection{Measured Noise}

To determine the effectiveness of the mosaicking process in reducing the \ac{RMS}, we performed a jack-knifing test for each GAMA field. We generated mosaics with varying numbers of included observations in a range from four to the maximum available, doubling the number of observations included each time. For each increment of the number of observations included, we created 100~mosaics with the included snapshot images being selected randomly from the total available. All observation snapshots used in the mosaics were resampled to the same grid.

The noise for each mosaic was estimated with \texttt{BANE}, measuring the average \ac{RMS} in each GAMA region. The results are shown in Figure~\ref{jackknife}, which shows the effect of a limiting noise on the mosaics. For both fields the noise decreases as expected theoretically (i.e. as the square-root of the number of observations), but then deviates with the co-addition of more than 16 snapshots for GAMA23 and more than 64 for GAMA9. Hence, the thermal noise dominates down to these levels, but after mosaicking more than 64--128 snapshot images, the decrease in RMS reduction tapers off considerably suggesting that the fields are reaching a confusion limit which is likely a combination of sidelobe and classical confusion. The lower decrease in RMS observed in the GAMA9 field was likely caused by a higher sidelobe confusion from the large number of bright sources in the field.

\begin{figure}[t!]
\centering
\includegraphics[width=0.98\linewidth]{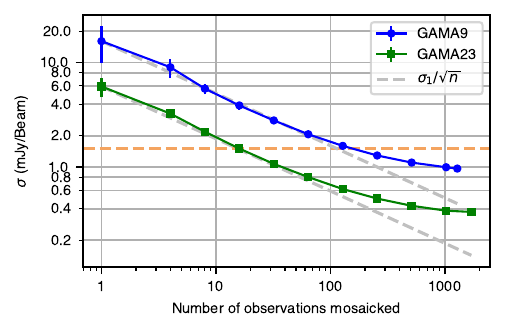}
\caption{Jackknife plot for the mean \ac{RMS} measured at the centre of the mosaic (y-axis) vs the number of observations mosaicked (x-axis). Each increment of the observation count was mosaicked 100~times with a randomised observation selection. The dashed grey lines represent the theoretical optimum noise reduction by the square-root of the number of observations. The dashed sandy brown line represents the median RMS achieved in GLEAM-X DRII (1.5\,mJy/beam). The \ac{RMS} reductions follows the theoretical optimum noise reduction for a mosaic count up to 64 snapshots for GAMA23 and 128 snapshots for GAMA9. After this point, as further snapshots are mosaicked, the RMS reduction tapers off indicating it has reached the sidelobe and/or classical confusion limit.}
\label{jackknife}
\end{figure}

Estimation of the final noise in the GAMA23 deep field mosaic is difficult because the image approaches the classical confusion limit. \texttt{BANE} does not naturally function in this regime as it expects sparse input images dominated by Gaussian thermal noise. For this reason, we used \texttt{BANE} to characterise the structure of the noise while implementing a novel approach to estimate the magnitude for the MIDAS mosaics.

We masked known sources in the $5^{\circ}\times 12^{\circ}$ GAMA23 region of our deep image using a higher frequency, deeper image produced from the EMU survey by:
\begin{itemize}
    \item Convolving the EMU deep field image of the \ac{GAMA}23 region at 887.5\,MHz produced by \textcite{gurkan2022} to the same beam size as the GAMA23 MIDAS mosaic.
    \item Re-projecting the convolved and the original EMU mosaics to the same pixel grid as the GAMA23 MIDAS mosaic.
    \item Masking pixels in the MIDAS GAMA23 mosaic where there was a corresponding pixel above the reported $5 \sigma$ threshold ($5 \times 0.038$\,mJy/beam) in either the original EMU mosaic, or the convolved EMU mosaic. This is equivalent to $5 \times 0.12$\,mJy/beam at the 215.68\,MHz observed in MIDAS assuming a spectral index of $\alpha=-0.8$.
    \item Dilating all masked pixels with a fully connected $3 \times 3$ structure function for two iterations using the \texttt{scipy} \texttt{ndimage} Python package.
\end{itemize}
This resulted in 49\% of the \ac{GAMA}23 region being masked with the remaining unmasked pixels in the region consisting mainly of noise. An example cutout of the implemented filter is presented in Appendix \ref{app_filter_fig} exhibiting its effectiveness in removing sources.

A histogram of the remaining pixels was plotted with a bin size of 5\,$\mu$Jy/beam and then a Gaussian fitted with \texttt{scipy} \texttt{curve\_fit} as seen in Figure~\ref{rms_hist} with the resultant values described in Table~\ref{table_fittedrms}. A Gaussian with a zero point $x_0=-96\pm 1$\,$\mu$Jy/beam and $\sigma = 331 \pm 1$\,$\mu$Jy/beam provided a very good fit with a small amount of additional positive flux observed due to incomplete masking. The masking of the bright areas leaves the underconvolved negative sidelobes of each bright source resulting in a $96$\,$\mu$Jy/beam offset to the zero point. This offset value is arbitrary and does not offer any meaningful insight. Therefore, we concluded the \ac{RMS} within the GAMA23 region is $0.331 \pm 0.001$\,mJy/beam.

\begin{table}[t!]
\begin{threeparttable}
\caption{The resultant values calculated in Section~\ref{rms} for a Gaussian fitted to the remaining pixel's values in the GAMA9 and GAMA23 mosaics after filtering out sources, used to estimate the RMS of the mosaics. The first column is the region mosaicked, the second column is the estimated classical confusion noise, the third is the theoretically predicted RMS calculated by Equation~\ref{Total_RMS_Equation}, the fourth is the $x_0$ value of the fitted Gaussian and the fifth is the reported $\sigma$.}
\label{table_fittedrms}
\begin{tabular}{lllll}
\toprule
\headrow  & \multicolumn{2}{c}{Predicted (mJy/beam)} & \multicolumn{2}{c}{Measured (mJy/beam)}\\
\headrow Mosaic & $\sigma_\mathrm{conf}$ & $\sigma_p$ & $x_0$ & $\sigma$ \\
\midrule
GAMA9 & $0.358$ & $0.833$ & $-0.171 \pm 0.001$ & $0.923 \pm 0.003$ \\ 
\midrule
GAMA23 & $0.231$ & $0.371$ & $-0.096 \pm 0.001$ & $0.331 \pm 0.001$ \\ 
\bottomrule
\end{tabular}
\end{threeparttable}
\end{table}

\begin{figure*}[t!]
\centering
\includegraphics[width=0.49\linewidth]{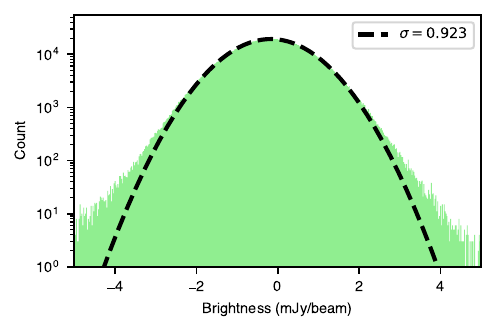}
\includegraphics[width=0.49\linewidth]{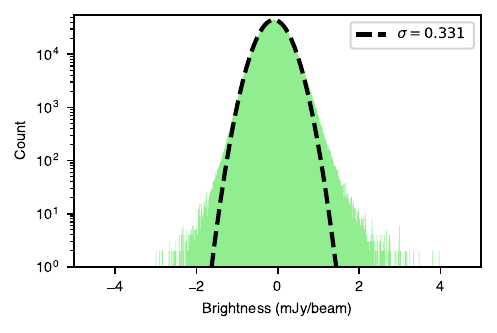}
\caption{Distribution of the pixel intensity in the GAMA9 region (left) and GAMA23 region (right) as a function of flux density for the deep images after masking all pixels above $5 \sigma$ in the EMU image. In the GAMA9 histogram, the black dashed line is a Gaussian with $x_0 = -0.171$\,mJy/beam and $\sigma = 0.923 $\,mJy/beam fit to the histogram. For the GAMA23 image, the black dashed line represents a Gaussian with $x_0 = -0.096$\,mJy/beam and $\sigma = 0.331$\,mJy/beam.}
\label{rms_hist}
\end{figure*}

To map the structure of the noise in the GAMA23 mosaic, \texttt{BANE} was run on the deep image with the default arguments. We then applied the same mask used to measure the RMS in the deep image to the \ac{RMS} map output by \texttt{BANE}. The median value for the unmasked pixels in the GAMA23 region was then used to determine the scaling factor $f=\sigma/\tilde{i}$ where $\sigma$ is the \ac{RMS} determined from the Gaussian fit and $\tilde{i}$ is the median of the GAMA23 region in the masked \ac{RMS} map. The entire \ac{RMS} map was then multiplied by the scaling factor, calculated as $f=0.81$, to determine the \ac{RMS} distribution in the \ac{GAMA}23 deep field image.

Due to the lower elevation of the GAMA9 field, the resulting \ac{RMS} is higher with a larger combined natural and sidelobe confusion limit. In addition, the sidelobe noise is spatially correlated to the noise in the mosaic. The same methods that were used to calculate the \ac{RMS} for the GAMA23 mosaic were used for the GAMA9 deep image.

A higher frequency deep ASKAP image mosaic for this field has not been published; however, the project `Survey with ASKAP of GAMA-09+X-ray' (SWAG-X; \citealt{moss2020}) has produced several tiles of this region. To create the base image required for the mask, we obtained seven tiles around GAMA9 from the SWAG-X survey that were conducted with ASKAP at 887.5\,MHz. We then co-added the tiles with \texttt{SWARP} and estimated the \ac{RMS} with \texttt{BANE}, using the mean of the \ac{RMS} within the \ac{GAMA}9 deep field ($\sigma=0.054$\,mJy/beam) to set our threshold of $5\sigma$ for the pixel masking. The resultant mask covered 65\% of the GAMA9 deep field.

A Gaussian with $x_0 = -0.171 \pm 0.001$\,mJy/beam and $\sigma = 0.923  \pm 0.001$\,mJy/beam was fit to the histogram of the masked region. As was done with the \ac{GAMA}23 deep image, \texttt{BANE} was run on the GAMA9 image and the \ac{RMS} map was scaled accordingly with a calculated scale factor $f=0.89$. The resulting \ac{RMS} is significantly higher than the GAMA23 image due to the observing limitations of the field. The off-zenith observations result in a higher \ac{RMS}, then the addition of the bright sources in the sidelobes increases the \ac{RMS} even further while additionally introducing a noise structure to the mosaic. This constraint on the RMS is corroborated when examining the ratios between the GAMA fields for the average RMS of the individual snapshots (outlined in Table~\ref{table_obs}) and the ratio of the RMS for the final mosaics (outlined in Table~\ref{table_fittedrms}), where $\sigma_\mathrm{G9}/\sigma_\mathrm{G23} = 2.8$ for both.

Due to the attenuation of the primary beam, the \ac{RMS} increases significantly at the edges of each mosaic as seen in Figure~\ref{fig_rmsmaps}. In the GAMA23 deep field image, the RMS reaches close to 6\,mJy/beam at the edge while in the GAMA9 mosaic the \ac{RMS} increases to slightly over 10\,mJy/beam at the edge. However, due to the large field-of-view of the MWA, the RMS in the central $60\,\deg^2$ GAMA fields remain relatively uniform.

\begin{figure*}[t!]
\centering
\includegraphics[width=0.454\linewidth]{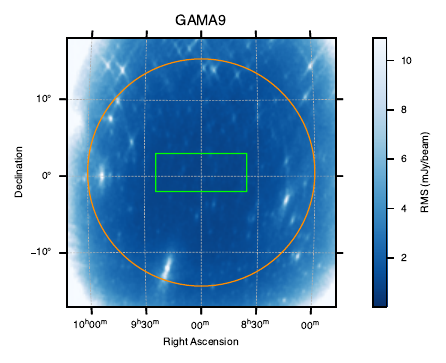}
\includegraphics[width=0.535\linewidth]{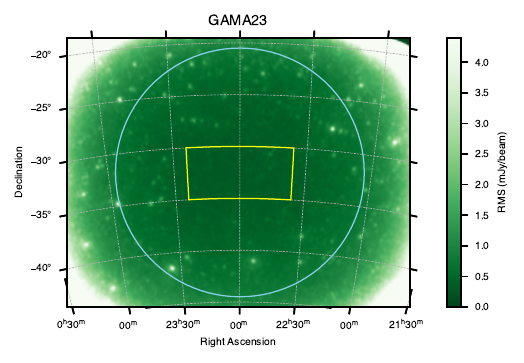}
\caption{RMS maps for the GAMA9 (left) and GAMA23 (right) MIDAS mosaics where the rectangles represent the regions targeted with GAMA9 in green and GAMA23 in yellow. The circles represent the area used for the completeness estimation and source counts for GAMA9 (orange) and GAMA23 (blue).}
\label{fig_rmsmaps}
\end{figure*}

\subsubsection{Theoretical Noise }

We estimated the theoretical classical confusion noise ($\sigma_\mathrm{conf}$) for each MIDAS mosaic using the methods derived by \textcite{condon1974} where

\begin{equation}
\label{Confusion_Equation}
\sigma_\mathrm{conf} = \left(\frac{q^{3-\gamma}}{3-\gamma}\right)^{\frac{1}{\gamma-1}} \left( k \frac{ \pi\,B_\mathrm{maj} B_\mathrm{min} }{4 (\gamma-1) \ln(2)}\right)^{\frac{1}{\gamma-1}}
\end{equation}

in which $B_\mathrm{maj}$ and $B_\mathrm{min}$ are the major and minor axes for the average restoring beam obtained by \textsc{WSClean} from the final fitted PSF listed in Table \ref{table_finalbeam} and $q$ is the \ac{SNR} that we take to be $q=1$ at the confusion limit. Using the LOFAR power-law source counts by \textcite{williams2016}, we estimate the normalisation $k=4 \times 10^{-6}$ with the source count slope $\gamma=1.5$ at flux densities around 1\,mJy. Note this slope has a conservatively steep value as the two fields only just reach the up-turn in the source counts at this flux density.

Combining the theoretical classical confusion noise with the theoretically optimal noise for $n$ number of observations mosaicked using the mean RMS from the individual snapshots ($\overline{\sigma_1}$) where
\begin{equation}
\label{Total_RMS_Equation}
\sigma_p = \frac{ ( \overline{\sigma_1} - \sigma_\mathrm{conf} ) }{ \sqrt{n} } + \sigma_\mathrm{conf}
\end{equation}
 resulted in the predicted RMS ($\sigma_p$) for the MIDAS mosaics listed in Table \ref{table_fittedrms}. However, this estimation does not account for sidelobe confusion that dominates GAMA9. Additionally, the confusion noise is highly sensitive to the source count slope which is rapidly changing at the faintest fluxes. We observe agreement with our predictions for GAMA23, while GAMA9 is slightly under-predicted as expected due to the large amount of sidelobe noise contributed by Hydra-A.

\begin{figure}[t!]
\centering
\includegraphics[width=0.98\linewidth]{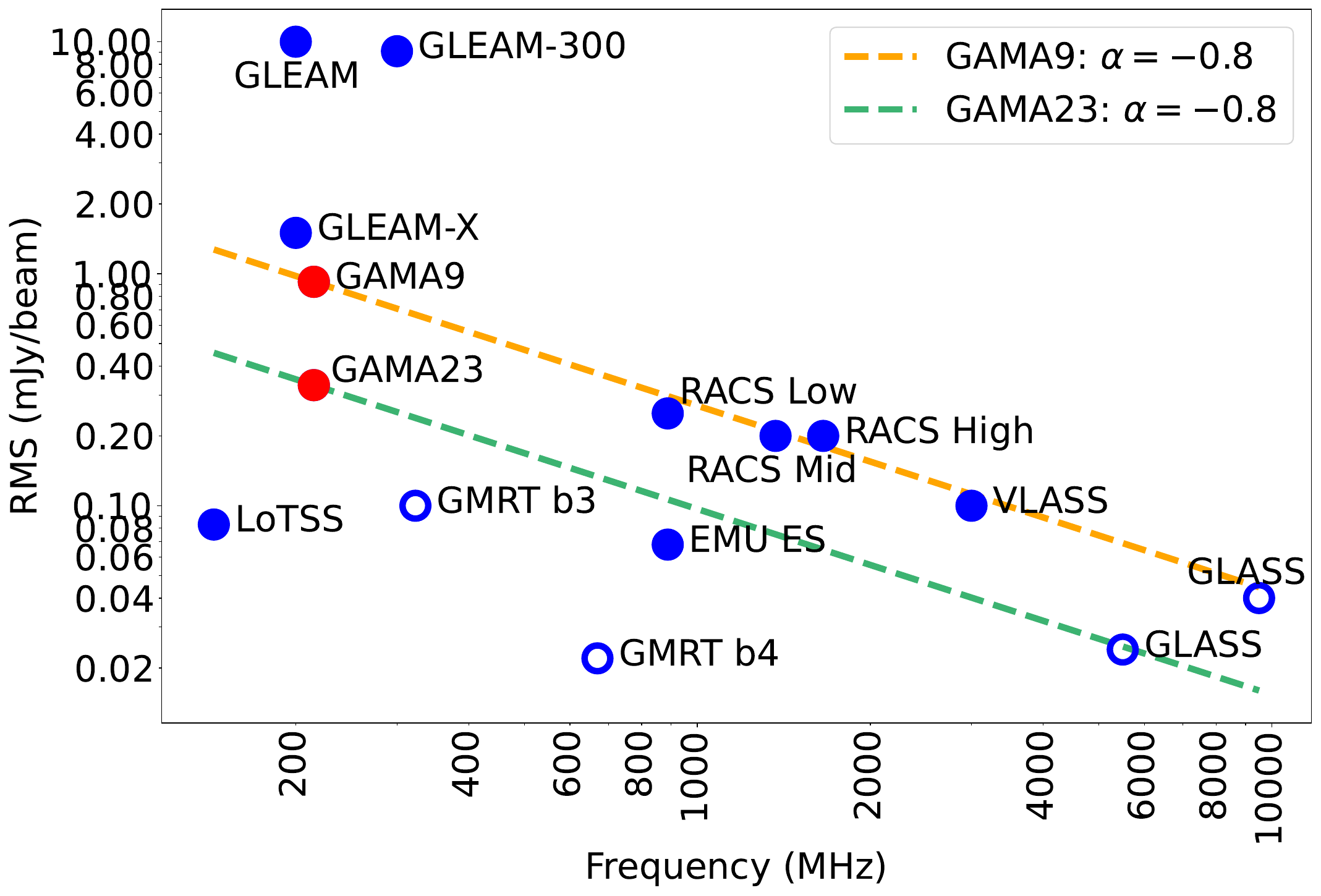}
\caption{Comparison of the 215.68\,MHz MIDAS GAMA9 and GAMA23 deep images to wide-field surveys ($>60 \text{\,deg}^2$) performed at radio wavelengths assuming a constant spectral index of $\alpha=-0.8$. Here the x-axis represents the wavelength of the survey measured in MHz and the y-axis reflects the \ac{RMS} of the deep field image measure in mJy/beam. The solid circles are completed published surveys while the open circles are estimations for unpublished surveys currently in progress.}
\label{fig_survey_comp}
\end{figure}

After scaling by a common spectral index ($\alpha=-0.8$ for $S_\nu\propto\nu^\alpha$) and comparing to wide-field ($>60 \text{\,deg}^2$) radio surveys in the Southern hemisphere, Figure~\ref{fig_survey_comp} shows that the GAMA9 mosaic reaches a depth deeper than GLEAM \autocite{hurley-walker2017, hurley-walker2019}, GLEAM-X \autocite{hurley-walker2022, ross2024} and GLEAM-300 \autocite{duchesne2025a}. It achieves an RMS comparable to surveys such as the Rapid ASKAP Continuum Survey (RACS-low; \citealt{mcconnell2020}, RACS-mid; \citealt{duchesne2023}, RACS-high; \citealt{duchesne2025}) and the Very Large Array Sky Survey (VLASS; \citealt{lacy2020}), probing a lower-frequency complementary range. However, the GAMA23 mosaic probes significantly deeper, offering a depth comparable to the expectant GAMA Legacy ATCA Sky Survey (GLASS; Huynh et al. in prep) and close to the 887.5\,MHz early EMU release \autocite{norris2021}. However, the MIDAS mosaics are much shallower than the LoTSS survey \autocite{shimwell2017, shimwell2019, shimwell2022} of the northern sky and the expected RMS for the Giant Metrewave Radio Telescope (GMRT; \citealt{seymour2020}) survey.

The MIDAS 215.68\,MHz deep field data along with the plethora of higher frequency data available for the \ac{GAMA}23 field offers an opportunity to probe this area of the sky further with the potential for new insight into galaxy clusters, halos, relics, \ac{AGN} and more.

\subsection{Source Catalogues} \label{sourcefinding}
The source catalogues for the MIDAS GAMA9 and GAMA23 surveys were created with \texttt{AEGEAN} using a blind source finding method. We used \texttt{AEGEAN}'s default parameters except for the \texttt{seedclip}, \ac{PSF}, \ac{RMS} and background values. The \texttt{seedclip} was set to $4 \sigma$ to detect pixels above this flux density for initial positions used in the component fitting. For the additional parameters, we utilised the \ac{PSF} generated in Section~\ref{blurring} along with the rescaled \ac{RMS} map and the constant background value calculated in Section~\ref{rms}. This resulted in a total of 67~985 sources found in the GAMA9 mosaic and 106~221 sources found in the GAMA23 mosaic, where we define a source as an individual fitted Gaussian component. The catalogues were reduced to sources with a peak flux density above the $5 \sigma$ detection level where the RMS is determined from the rescaled RMS maps produced in Section~\ref{rms}. Next, the sources within 1$'$ from the edge of the mosaic were removed from the GAMA catalogues to ensure that there were no sources with flux density overlapping the edge of the mosaic.

We visually inspected sources with high integrated to peak flux density ratios in both MIDAS mosaics. It was determined all sources with a ratio $S_\mathrm{int}/S_\mathrm{peak} > 16$ were spurious and therefore removed from the catalogues. Due to the higher noise around bright sources caused by sidelobes, sources nearby are difficult to validate as they could be artefacts from the sidelobes. As a precaution, we flagged sources that were within $4'$ of a 1.5\,Jy sources as lower reliability. As a result, 429 sources in the GAMA9 catalogue and 852 sources in the GAMA23 catalogue were flagged.

As detailed in Table~\ref{table_catsources}, this resulted in a total of 53~419 sources in the final GAMA9 catalogue and 80~407 sources in the GAMA23 catalogue. The 24 column names in the MIDAS catalogues are outlined in Appendix \ref{app_colnames}. For GAMA9, 3.4\% of islands were multi-component while 5.0\% of the islands in GAMA23 had multiple components.

\begin{table}[t!]
\begin{threeparttable}
\caption{The number of sources detected within each survey, the rows represent each cut that was applied. Each column lists the numbers of sources remaining after each cut where the full image columns are the number of sources within the full survey mosaic and the region columns are the number of sources within the GAMA9 (RA $+129^{\circ}$ to $+141^{\circ}$, Dec $-2^{\circ}$ to $+3^{\circ}$) and GAMA23 (RA $+339^{\circ}$ to $+351^{\circ}$, Dec $-35^{\circ}$ to $-30^{\circ}$) regions respectively.}
\label{table_catsources}
\begin{tabular}{lllll}
\toprule
\headrow  & \multicolumn{2}{c}{GAMA9} & \multicolumn{2}{c}{GAMA23} \\
\headrow Cut & Full Image & Region & Full Image & Region \\
\midrule
Initial & 67~985 & 5~043 & 106~221 & 10~591 \\
\midrule
$S_\mathrm{peak} < 5 \sigma$ & 53~515 & 4~227 & 80~523 & 8~423 \\
\midrule
On Edge & 53~431 & 4~227 & 80~428 & 8~423 \\
\midrule
$S_\mathrm{int}/S_\mathrm{peak} > 16$ & 53~419 & 4~225 & 80~407 & 8~417 \\
\bottomrule
\end{tabular}
\end{threeparttable}
\end{table}
}

The average beam dimensions were calculated from the source dimensions of compact sources within the MIDAS catalogues selected with the RMS envelope described in Section~\ref{fluxscalecomparison}. The average $B_\mathrm{maj}$ and $B_\mathrm{min}$ for sources within the respective GAMA regions were calculated and are outlined in Table~\ref{table_finalbeam}. Within the GAMA9 region, 4~225 sources were detected with a beam size of $2.73 \times 10^{-4} \text{\,deg}^2$ for the mosaic, resulting in a source density of 70.42\,$\text{sources} / \text{deg}^2$ (52.0 beams per source\footnote{This is a classical measure of how dense the field is, where 30 and less is a typical value for classical confusion.}). However, a much larger number of sources were detected within the GAMA23 region with a total of 8~417 sources. With a beam size of $2.20 \times 10^{-4} \text{\,deg}^2$ for the mosaic, we approach the classical confusion limit with a source density of 140.28\,$\text{sources} / \text{deg}^2$ (32.4 beams per source).

\begin{table*}[t!]
\begin{threeparttable}
\caption{The beam dimensions calculated from the source dimensions of unresolved sources in the MIDAS catalogues along with the resultant beams per source calculated for each GAMA field's mosaic.}
\label{table_finalbeam}
\begin{tabular}{llllll}
\toprule
\headrow Field & $B_\mathrm{maj}$ (arcsec) & $B_\mathrm{min}$ (arcsec) & $B_\mathrm{pa}$ (degrees) & Beam Area ($\text{deg}^2$) & Beams per Source \\
\midrule
GAMA9 & $64.3$ & $48.7$ & $-66.1$ & $2.73 \times 10^{-4}$ & $52.0$ \\
\midrule
GAMA23 & $57.3$ & $43.9$ & $-27.4$ & $2.20 \times 10^{-4}$ & $32.4$ \\
\bottomrule
\end{tabular}
\end{threeparttable}
\end{table*}

\subsection {Flux Density Scale Evaluation}\label{fluxscalecomparison}

To evaluate the flux density scales of our catalogues, we compared the MIDAS catalogues to the wideband integrated flux densities from the GLEAM catalogue \autocite{hurley-walker2017} and to the GLEAM-X DRII catalogue \autocite{ross2024} for GAMA23. The GLEAM(-X) flux densities were scaled to 215.68\,MHz using an assumed constant spectral index of $\alpha = -0.8$. At the time of writing, there are no public data releases from GLEAM-X that cover the GAMA9 field.

\begin{figure*}[t!]
\centering
\includegraphics[width=0.49\linewidth]{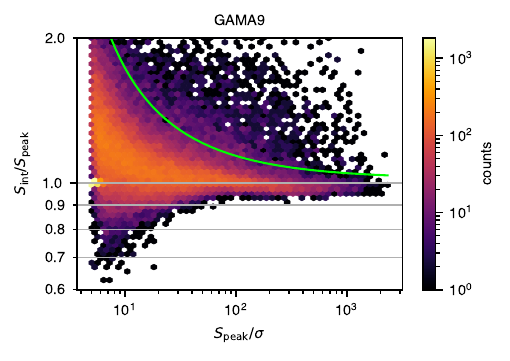}
\includegraphics[width=0.49\linewidth]{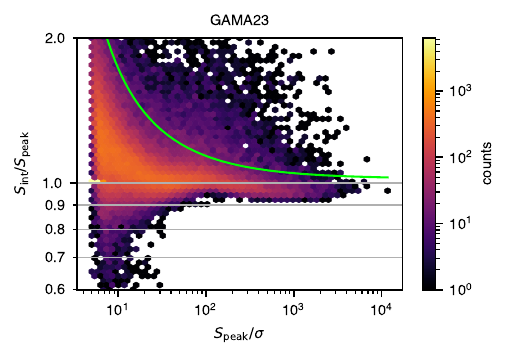}
\caption{Ratio of integrated flux density to peak flux density as a function of SNR for the MIDAS GAMA9 (left) and GAMA23 (right) source catalogues. The green line is the upper envelope, equation~\ref{rms-envelope}, set to define the limit for unresolved sources that we approximate to be 95\% of the total sources. The small bump observed in GAMA23 around $S_\mathrm{peak}/\sigma \approx 100$ and the less noticeable bump observed in GAMA9 around $S_\mathrm{peak}/\sigma \approx 60$ is suspected to be a result of sources on the detection limit in the individual observations; these sources would be deconvolved in some observations, but not in others.}
\label{int_peak}
\end{figure*}

The MIDAS catalogues were matched to GLEAM(-X) with a maximum separation distance of $1'$. The MIDAS catalogue was reduced to only include compact source, additionally any MIDAS source with a neighbour within $2'$ was removed to ensure that sources within the GLEAM(-X) catalogues were not resolved into multiple sources within the MIDAS catalogues. Compact sources were selected from the catalogue based on their integrated flux density to peak flux density ratio. We plotted the ratios as a function of SNR as seen in Figure~\ref{int_peak} where we estimated an upper envelope of

\begin{equation}
\label{rms-envelope}
\frac{S_\mathrm{peak}}{S_\mathrm{int}} = 0.98 - \frac{1.6}{(S_\mathrm{peak}/\sigma)^{0.6}}
\end{equation}

to capture 95\% of the sources, where $S_\mathrm{peak}$ is the measured peak flux, $S_\mathrm{int}$ is the measured integrated flux and $\sigma$ is the measured local RMS. A small bump is observed in the ratios around $S_\mathrm{peak}/\sigma \approx 60$--$100$ that is likely caused by un\texttt{CLEAN}ed sources. The masking stage discussed in Section~\ref{Masking} results in only sources $>3\sigma_i$ being \texttt{CLEAN}ed (where $\sigma_i$ is the local RMS in an individual 2-min snapshot). Therefore, sources with a flux density near this detection limit will be \texttt{CLEAN}ed in some snapshots and not others, resulting in more unpredictable source morphologies after mosaicking. Sources below this detection limit will not be \texttt{CLEAN}ed at all.

A correctable bias in the flux density measurements can be introduced from measurement errors known as Eddington bias \autocite{eddington1913}. Flux density measurements for sources will scatter above and below their true values. As there are a greater number of the fainter sources (due to the steep source counts), that will be measured brighter than there are bright sources that will be measured fainter, the flux density measurements for the faint sources are much more likely to be overestimated. To evaluate the Eddington bias in our flux density scale comparisons we use the maximum-likelihood value ($S_\mathrm{ml}$; \citealt{hogg1998, hurley-walker2017}),

\begin{equation}
\label{eddington-bias-correction}
S_\mathrm{ml} = \frac{S}{2} + \frac{S}{2} \sqrt{1 - \frac{4q+4}{(S / \sigma)^2}}
\end{equation}

where $S$ is the measured integrated flux density, $\sigma$ is the local RMS measured for the source and $q=1.54$ is the logarithmic source count slope at the GLEAM flux densities \autocite{franzen2016,hurley-walker2022,ross2024}. This correction made a negligible difference in our results so we chose not to apply it.

\begin{figure*}[t!]
\centering
\includegraphics[width=1.0\linewidth]{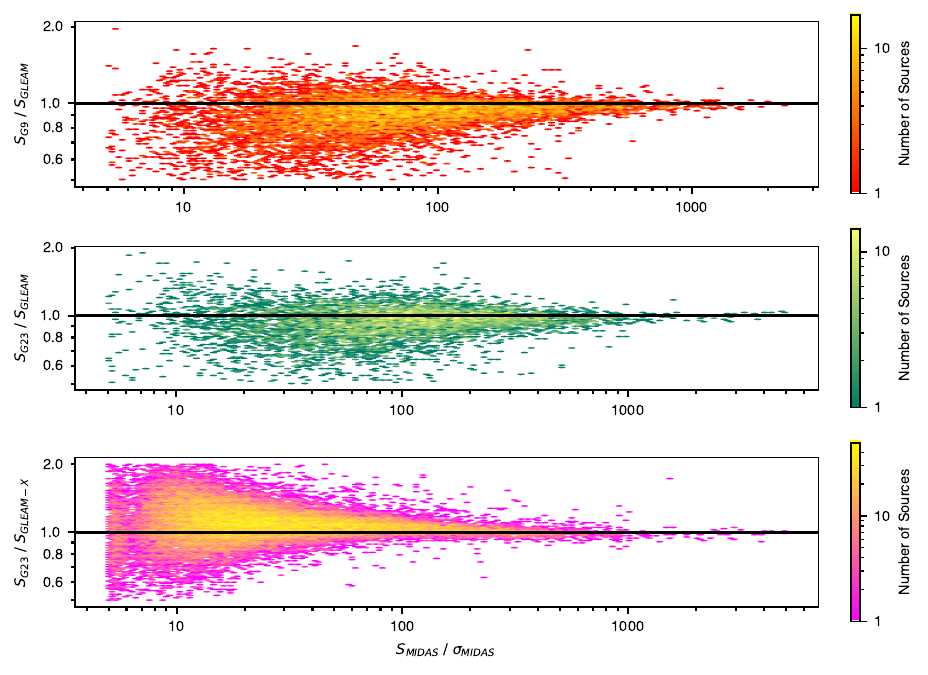}
\caption{The integrated flux density ratios between GLEAM and GAMA9 MIDAS at the top, GLEAM and GAMA23 MIDAS in the middle and GLEAM-X and GAMA23 MIDAS at the bottom. The x-axes represents the MIDAS \ac{SNR} and the y-axes is the flux density ratio between the surveys. The solid black line represent a ratio of 1. Here we observe good agreement with GLEAM through the entire flux density range. We observe agreement with GLEAM-X at an SNR $> {\sim} 100$, however at lower SNR values GLEAM-X reports lower flux density scales. We attribute this to the \texttt{CLEAN}ing methods used in GLEAM-X resulting in an over-estimate of the observed vs true flux density values.}
\label{fluxscale_comp}
\end{figure*}

The flux density scale comparisons with GLEAM(-X) in Figure~\ref{fluxscale_comp} exhibits good agreement with GLEAM. For GLEAM-X DRII we observe agreeance at an SNR $> {\sim} 100$, however at lower SNR values we observe a discrepancy where GLEAM-X reports lower source flux densities. This is likely due to the \texttt{CLEAN}ing methods used in GLEAM-X where a $3\sigma$ \texttt{auto-mask} was used. As discussed in Section \ref{Masking}, we found this level of \texttt{auto-mask} would \texttt{CLEAN} noise peaks removing flux from real sources and introduce a small offset into the flux density measurements. This results in an under-prediction of flux density which is more evident for fainter sources and aligns with the observed discrepancy.

The flux density scaling for MIDAS was performed using GLEAM as a reference during post-imaging as described in Section~\ref{post-image}. GLEAM reports an uncertainty of 8\% for its flux density scale accuracy, therefore we prescribe an 8\% uncertainty for the MIDAS flux density scaling when compared to other surveys.

\subsection {Astrometry}
The precision of the astrometry for the final MIDAS GAMA9 and GAMA23 source catalogues was evaluated by comparing the source positions with the joint NVSS and SUMSS positional-based catalogue of unresolved sources described in Section 3.3, which we will refer to as the reference catalogue. The GAMA deep field catalogues were reduced to sources $>50\sigma$ and matched to the reference catalogue with a matching distance within $1'$, where the source positions in the reference catalogue were assumed to be the correct positions. We then calculated the offsets for the deep field sources relative to the reference sources and calculate the mean offset and standard deviation for the offset per catalogue.

\begin{figure*}[t!]
\centering
\includegraphics[width=0.49\linewidth]{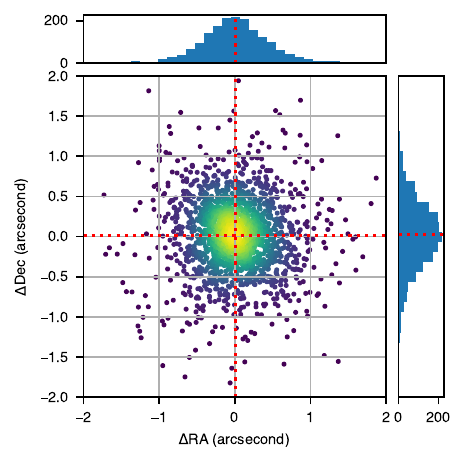}
\includegraphics[width=0.49\linewidth]{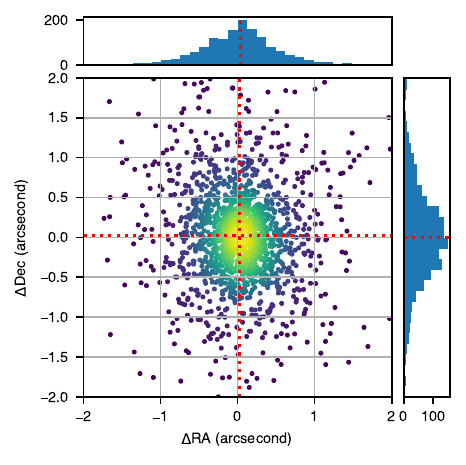}
\caption{The astrometric offsets for the GAMA9 source catalogue (left) and GAMA23 source catalogue (right) compared to a reference catalogue containing
unresolved sources from NVSS and SUMSS. The x-axes and y-axes are the offsets for the RA and Dec respectively measured in arcseconds. A total off 8~749 sources was compared resulting in an RA offset of $+0.007'' \pm 0.706''$ and a Dec offset of $+0.002'' \pm 1.440''$ for GAMA9 and a total 13~355 sources for GAMA23 resulting in an average RA offset of $-0.009'' \pm 1.054''$ and a Dec offset of $+0.010'' \pm 1.149''$. The red dotted lines represents the mean RA and Dec offsets.}
\label{astrometry}
\end{figure*}

For the GAMA9 catalogue, there were a total of 8~749 sources above $50\sigma$ matched to the reference catalogue. The resulting mean RA offset was $+0.007'' \pm 0.706''$ and the mean Dec offset was $+0.002'' \pm 1.440''$ as seen in Figure~\ref{astrometry}.

The GAMA23 catalogue matched 13~355 sources above $50\sigma$ to the reference catalogue. As seen in Figure~\ref{astrometry}, an average RA offset of $-0.009'' \pm 1.054''$ and a Dec offset of $+0.010'' \pm 1.149''$ was calculated.

We additionally confirm astrometry matching to RACS-high which offers a higher angular resolution of $11.8'' \times 8.1''$ but experiences astrometry limitations with positional uncertainties of $0.6''$ in right-ascension and $0.7''$ in declination \autocite{duchesne2025}. An average RA offset of $+0.040''\pm3.916''$ and Dec offset of $-0.095''\pm3.666''$ was calculated between the GAMA9 catalogue and the RACS-high catalogue. For the GAMA23 catalogue, an average RA offset of $+0.24''\pm3.97''$ and a Dec offset of $-0.091''\pm3.904''$\ was calculated. Due to the limitations of RACS-high, the results have large uncertainties; however, they are in good agreement.

\subsection {Completeness}\label{completeness}
To estimate the percentage of true sources recovered at a given flux density, which we define as the completeness, we injected simulated sources into random positions in each deep field image at selected flux density levels. To accomplish this, we used \texttt{AeRes} from the \texttt{AEGEAN} package that enables adding or removing sources from an image with specified positions, flux densities and shapes. We ran 26 simulations for flux density levels ranging from 1\,mJy to 316\,mJy, evenly spaced in log space, injecting ${\sim}10$ point-like sources per square-degree at a constant flux density for each simulation\footnote{Formally the number of sources injected per flux density bin should vary per the source counts to obtain the most precise uncertainties. This was verified with a small number of simulations which confirmed that injecting more sources resulted in a consistent completeness, but reduced the uncertainties. However, this is compute intensive and testing demonstrated that the completeness uncertainties were always far less than the Poisson uncertainties so the extra precision was unnecessary.}. To ensure that we were not introducing an unnecessary confusion measurement, a minimum separation distance between simulated sources was set at 5$'$. No minimum separation distance was established between the simulated and real sources to ensure that the classical confusion limit was accurately taken into account.

Blind source finding with \texttt{AEGEAN} was run on each simulation and the percentage of injected sources recovered was calculated. To estimate the uncertainty for the completeness calculations, we ran the full simulation process 100 times for each region. The mean for each flux density level was then used for our completeness values and the standard deviation for the uncertainty. 

\begin{figure}[t!]
\centering
\includegraphics[width=0.98\linewidth]{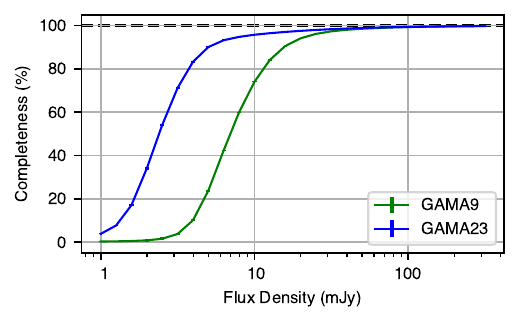}
\caption{The completeness for a circular area with a radius of $15^{\circ}$ in the GAMA9 (green) mosaic and a circle with a radius of $12^{\circ}$ in the GAMA23 (blue) mosaic as a function of flux density. The errors were calculated as the standard deviation from 100 simulations. The dashed black line represents 100\% completeness. A 90\% completeness was estimated at ${\sim}10.06$\,mJy for GAMA9 and ${\sim}4.97$\,mJy for GAMA23.}
\label{fig_completeness}
\end{figure}

The completeness curves for the GAMA regions are shown in Figure~\ref{fig_completeness}, although part of the shape of the curves will be due to the increase in the RMS towards the edges of the fields of view that is observed in Figure \ref{fig_rmsmaps}. To attempt to minimise this effect, we measure the completeness for a circular area in the centre of each region. For the GAMA9 field, a circular area with a radius of $15^{\circ}$ was used resulting in an estimated completeness of 50\% at ${\sim}7.0$\,mJy and 90\% at ${\sim}15.6$\,mJy. A $12^{\circ}$ radius was used for the circle at the centre of the GAMA23 region resulting in a completeness of 50\% at ${\sim}2.4$\,mJy and 90\% at ${\sim}5.0$\,mJy.

\subsection {Source Counts}\label{sourceounts}

The source counts for each deep field catalogue were calculated for a circular area at the centre of each GAMA region. For the GAMA9 mosaic, we used a circle with a radius of $15^{\circ}$ centred at RA $+135^{\circ}$, Dec $+0.5^{\circ}$; for the GAMA23 mosaic, we used a circle with a radius of $12^{\circ}$ centred at RA $+345^{\circ}$, Dec $-32.5^{\circ}$ (i.e. the same regions over which we determined the completeness). The source Hydra-A was excluded from the GAMA9 MIDAS source counts as it is a very bright outlier. The sources detected in each circular region were binned by flux density level in 15 bins; the faintest 12 bins are equal size in log space, while the three brightest bins were double the size to account for the lower number of sources. The value for the start of the first bin was calculated as five times the average \ac{RMS} of all selected sources, while the end of the final bin was set as the value of the brightest source.

The GAMA9 source catalogue was binned from $7.44 \times 10^{-3}$\,Jy to 23.3\,Jy with a bin width of $\Delta log_{10}(S) = 0.194$\,Jy for the first 12 bins and a width of $\Delta log_{10}(S) = 0.388$\,Jy for the brightest 3 bins. For GAMA23 we used a bin width of $\Delta log_{10}(S) = 0.217$ in log-space for the first 24 bins and $\Delta log_{10}(S) = 0.434$\,Jy for the 3 brightest bins, covering a flux density range from $2.58 \times 10^{-3}$\,Jy to 21.0\,Jy.

\begin{figure*}[t!]
\centering
\includegraphics[width=1.0\linewidth]{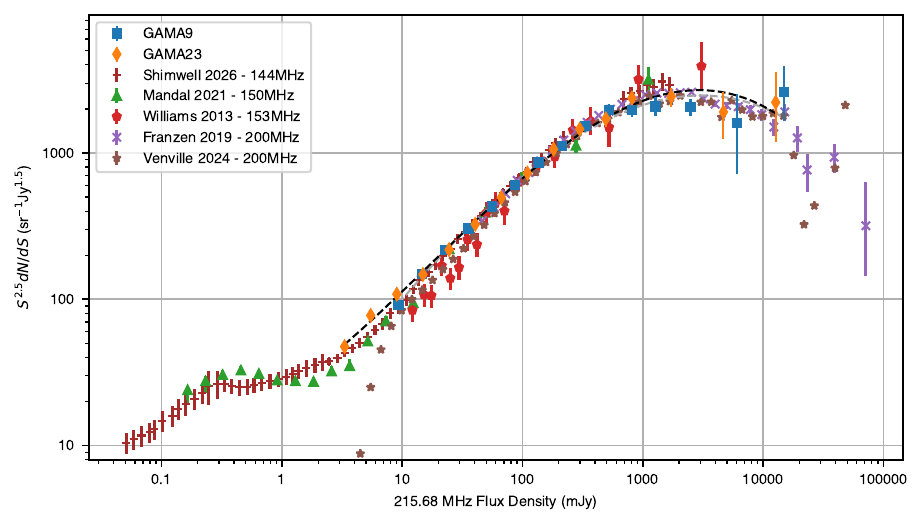}
\caption{Source counts normalised to Euclidean values for GAMA9 (blue) and GAMA23 (gold) as a function of flux density in comparison to low-frequency counts from literature scaled to 215.68\,MHz. Here we observe good agreement with the 200\,MHz counts by \textcite{franzen2019}, but a minor deviation from \textcite{venville2024} at a flux density below ${\sim} 40$\,mJy. We attribute this deviation to the lack of a completeness correction applied to their source counts. We observe good agreement with the 144\,MHz source counts by \textcite{shimwell2026} with a slight deviation at the bright end with GAMA9 and a small deviation at the faint end with GAMA23 as it approaches the GAMA9 observing limit. Comparing to the ${\sim}150$\,MHz counts by \textcite{mandal2021} and \textcite{williams2013}, we observe agreeance down to ${\sim} 25$\,mJy. Below this we observe a deviation in the counts that is likely due to the differing resolutions and flux density limits of the surveys along with the different source finding methods employed. The black and grey dashed lines represent a fitted third-degree polynomial to the GAMA23 and GAMA9 source counts respectively.}
\label{fig_sourcecounts}
\end{figure*}

\begin{table}[t!]
\begin{threeparttable}
\caption{The source counts calculated with the GAMA9 mosaic. The first column is the bin ranges, the second is the bin centres and the third is the number of sources detected in the bin. The fourth column is the correction factor applied to the number of sources, calculated from the completeness presented in Section~\ref{completeness}. The fifth column is the source counts normalised to Euclidean values along with their uncertainty estimated from Poisson statistics, the uncertainty due to cosmic variance and completeness errors added in quadrature.}
\label{table_sourcecounts_g9}
\begin{tabular}{ccccc}
\toprule
\headrow Bin Range & Bin Centre & & Correction & $S_c^{2.5} dN/dS$ \\
\headrow $S$ (mJy) & $S$ (mJy) & $N_s$ & Factor & (sr${}^{-1}$Jy${}^{1.5}$) \\
\midrule
$7.44$ -- $11.6$ & $9.30$ & $6809$ & $1.444\pm0.004$ & $91.4^{+5.7}_{-5.7}$ \\
\midrule
$11.6$ -- $18.2$ & $14.54$ & $7218$ & $1.137\pm0.003$ & $149^{+8.5}_{-8.5}$ \\
\midrule
$18.2$ -- $28.4$ & $22.75$ & $5835$ & $1.050\pm0.002$ & $218^{+12}_{-12}$ \\
\midrule
$28.4$ -- $44.5$ & $35.58$ & $4281$ & $1.024\pm0.002$ & $305^{+17}_{-17}$ \\
\midrule
$44.5$ -- $69.6$ & $55.64$ & $3121$ & $1.013\pm0.001$ & $430^{+25}_{-25}$ \\
\midrule
$69.6$ -- $109$ & $87.02$ & $2244$ & $1.008\pm0.001$ & $601^{+39}_{-39}$ \\
\midrule
$109$ -- $170$ & $136.1$ & $1642$ & $1.005\pm0.001$ & $858^{+67}_{-66}$ \\
\midrule
$170$ -- $266$ & $212.8$ & $1095$ & $1.003\pm0.001$ & $1~117^{+86}_{-86}$ \\
\midrule
$266$ -- $416$ & $332.9$ & $767$ & -- & $1~526^{+132}_{-131}$ \\
\midrule
$416$ -- $651$ & $520.6$ & $504$ & -- & $1~961^{+188}_{-187}$ \\
\midrule
$651$ -- $1~018$ & $814.2$ & $259$ & -- & $1~971^{+209}_{-204}$ \\
\midrule
$1~018$ -- $1~592$ & $1~273$ & $140$ & -- & $2~084^{+317}_{-307}$ \\
\midrule
$1~592$ -- $3~895$ & $2~490$ & $104$ & -- & $2~065^{+281}_{-264}$ \\
\midrule
$3~895$ -- $9~527$ & $6~092$ & $21$ & -- & $1~595^{+914}_{-877}$ \\
\midrule
$9~527$ -- $23~303$ & $14~900$ & $9$ & -- & $2~615^{+1~293}_{-973}$ \\
\bottomrule
\end{tabular}
\end{threeparttable}
\end{table}

\begin{table}[t!]
\begin{threeparttable}
\caption{The source counts calculated with the GAMA23 mosaic. As with Table~\ref{table_sourcecounts_g9}, The first column is the bin ranges, the second is the bin centres and the third is the number of sources detected in the bin. The fourth column is the correction factor applied to the number of sources, calculated from the completeness calculated in Section~\ref{completeness}. The fifth column is the source count normalised to Euclidean values along with their uncertainty estimated from Poisson statistics, the uncertainty due to cosmic variance and completeness errors added in quadrature.}
\label{table_sourcecounts_g23}
\begin{tabular}{ccccc}
\toprule
\headrow Bin Range & Bin Centre & & Correction & $S_c^{2.5} dN/dS$ \\
\headrow $S$ (mJy) & $S$ (mJy) & $N_s$ & Factor & (sr${}^{-1}$Jy${}^{1.5}$) \\
\midrule
$2.58$ -- $4.26$ & $3.32$ & $12654$ & $1.361\pm0.005$ & $47.4^{+2.9}_{-2.9}$ \\
\midrule
$4.26$ -- $7.02$ & $5.47$ & $12115$ & $1.097\pm0.004$ & $77.4^{+3.9}_{-3.9}$ \\
\midrule
$7.02$ -- $11.6$ & $9.02$ & $8354$ & $1.050\pm0.003$ & $108^{+6.2}_{-6.2}$ \\
\midrule
$11.6$ -- $19.1$ & $14.87$ & $5468$ & $1.032\pm0.003$ & $147^{+9.5}_{-9.5}$ \\
\midrule
$19.1$ -- $31.5$ & $24.51$ & $3856$ & $1.021\pm0.002$ & $218^{+14}_{-14}$ \\
\midrule
$31.5$ -- $51.9$ & $40.42$ & $2740$ & $1.014\pm0.002$ & $325^{+22}_{-21}$ \\
\midrule
$51.9$ -- $85.6$ & $66.64$ & $1978$ & $1.009\pm0.001$ & $495^{+36}_{-36}$ \\
\midrule
$85.6$ -- $141$ & $109.9$ & $1378$ & $1.006\pm0.001$ & $728^{+54}_{-53}$ \\
\midrule
$141$ -- $233$ & $181.2$ & $949$ & $1.004\pm0.001$ & $1~059^{+86}_{-85}$ \\
\midrule
$233$ -- $384$ & $298.8$ & $616$ & $1.003\pm0.001$ & $1~454^{+135}_{-134}$ \\
\midrule
$384$ -- $633$ & $492.6$ & $343$ & -- & $1~709^{+202}_{-200}$ \\
\midrule
$633$ -- $1~043$ & $812.2$ & $225$ & -- & $2~374^{+255}_{-248}$ \\
\midrule
$1~043$ -- $2~835$ & $1~720$ & $154$ & -- & $2~426^{+368}_{-359}$ \\
\midrule
$2~835$ -- $7~708$ & $4~675$ & $27$ & -- & $1~907^{+710}_{-663}$ \\
\midrule
$7~708$ -- $20~956$ & $12~710$ & $7$ & -- & $2~216^{+1~359}_{-1~026}$ \\
\bottomrule
\end{tabular}
\end{threeparttable}
\end{table}

Here in Figure~\ref{fig_sourcecounts}, we present the source counts for our deep field catalogues normalised to Euclidean values. These source counts have been corrected for the incompleteness calculated in Section~\ref{completeness} and are listed in Tables~\ref{table_sourcecounts_g9}~\&~\ref{table_sourcecounts_g23}. The corrected normalised source counts ($C_N$) were calculated by

\begin{equation}
\label{Sourcecount_Equation}
C_N = S^{2.5} \frac{dN}{a\,dS}
\end{equation}

where $S$ is the centre of the flux density bins in Jy, $dN$ is the number of sources in the bin ($N_s$) multiplied by the correction factor for completeness calculated from the linear fit between the two closest flux density bins determined in Section~\ref{completeness}, $a$ is the area sampled in steradian and $dS$ is the width of the bin in Jy.

The uncertainty for the source counts was estimated using Poisson statistics for a confidence interval equivalent to a $1 \sigma$ Gaussian described by \textcite{gehrels1986} where for the upper ($\lambda_u$) and lower ($\lambda_l$) limits,

\begin{equation}
\label{Sourcecount_Error_Equation}
\begin{split}
\lambda_u \approx \sqrt{dN + 1} + 1 \\
\lambda_l \approx \sqrt{dN - 0.25}
\end{split}.
\end{equation}

For large values of $dN$ these uncertainties tend towards $\lambda=\sqrt{dN}$, however these uncertainties are formally correct for $dN\lesssim 20$.

Additionally, we add cosmic variance to the uncertainty, which we calculate for each bin using the \ac{LoTSS} Data Release 2 catalogue \autocite{shimwell2022}. The data release specifies a potential flux density variation dependant on pointing which may limit this analysis, but we continue with the assumption the effects are small. The flux density of these sources was scaled from 144\,MHz to 215.68\,MHz under the assumption of a constant spectral index of $\alpha = -0.8$. The source counts for each bin in each region were measured for 100~randomly positioned circular areas within the area surveyed by LoTSS DR2. The size of the circles were equal to those that were used to measure the deep field GAMA source counts, $15^{\circ}$ for GAMA9 and $12^{\circ}$ for GAMA23. The centre location of the circles were determined by selecting random sources in the \ac{LoTSS} catalogue. To ensure that the region was not on the edge of the measured field and had complete coverage, we searched an area of 2$'$ at the upper and lower extremes for the RA and Dec coordinates of the circle's edges to ensure there were sources present. The cosmic variance for the flux density bins below 2\,Jy was calculated to be ${\sim} 5$--$10$\%, but for the brighter bins the cosmic variance increases significantly to ${\sim} 20$--$35$\%. The error on the corrected source counts was taken as the sum in quadrature of the Poisson error, completeness correction error, and uncertainty due to cosmic variance.

The resulting normalised source counts seen in Figure~\ref{fig_sourcecounts} are compared to the counts generated from other low-frequency surveys scaled by $\alpha=-0.8$. We can not compare source counts from surveys at very different frequencies due to selection biases resulting in changes to spectral index distribution between the frequency ranges. This means a simple spectral index conversion could not be used for comparison with surveys completed at frequencies significantly different from the 215.68\,MHz we observed at \autocite{kellermann1964, franzen2016}. Unfortunately, no source counts have been previously calculated at 216\,MHz, so we made comparisons with surveys conducted at 144\,MHz, 150\,MHz, 153\,MHz and 200\,MHz.

For our comparisons, we used counts by \textcite{williams2013} who utilised observations of the National Optical Astronomy Observatory (NOAO) Boötes field with the Giant Metrewave Radio Telescope (GMRT) at 153\,MHz producing an image reaching an RMS of $2$\,mJy/beam covering an area of $30 \text{\,deg}^2$. Additionally, we included source counts at 200\,MHz from \textcite{franzen2019} using the GLEAM survey that reaches an RMS of $10$\,mJy/beam and covers an area of $25~000\text{\,deg}^2$, which reduces potential influence of cosmic variance and improves accuracy at the bright end. We also include the 200\,MHz source counts produced by \textcite{venville2024} that utilises the central $5~600 \text{\,deg}^2$ area of the GLEAM-X DR2 survey which contains the lowest and least variable RMS properties but do not correct for completeness. Finally, we included source counts produced from the LoTSS survey which offer better insight at the faint end. We include the 150\,MHz counts by \textcite{mandal2021} that utilises the deep field observations from the survey reaching an RMS of 0.1\,mJy/beam for a $25\text{\,deg}^2$ area along with the counts published in LoTSS-DR3 by \textcite{shimwell2026} covering 88\% of the northern sky, achieving an average RMS of 0.092\,mJy/beam.

Comparing to the 200\,MHz counts, we see good agreement with \textcite{franzen2016} for the entire flux density range, reaffirming the alignment with GLEAM as observed in Section~\ref{fluxscalecomparison}. The comparison with \textcite{venville2024} shows good agreement down to ${\sim} 40$\,mJy, below this level their source count slope becomes steeper. This is consistent with the lack of a completeness correction applied to their source counts, where the GLEAM-X completeness estimations exhibit a decrease in this range \autocite{ross2024}. Additionally, the under-predicated flux densities in GLEAM-X for fainter sources as discussed in Section~\ref{Masking} and Section~\ref{fluxscalecomparison} will have additional effects on the reported source counts at lower flux densities.

For the comparisons with the 144\,MHz source counts by \textcite{shimwell2026}, we see good agreeance for GAMA9 with a slight deviation at the bright end and good agreement with GAMA23 with a small deviation at flux densities $<10$\,mJy/beam where we approach the observing limit of GAMA9. The comparisons with the ${\sim}150$\,MHz source counts, exhibit agreement with \textcite{mandal2021} down to ${\sim} 25$\,mJy. \textcite{williams2013} shows increasing scatter around these flux density levels, but we observe agreeance with their higher calculations. Below this point we see a deviation in the counts where several factors are likely at influence. Cosmic variance due to the smaller survey area may be a contributing factor. Additionally, the source counts are based on surveys with a higher resolution that likely resolves some sources into multiple fainter sources instead of a single brighter source observed in MIDAS. Finally, the MIDAS source catalogues were produced with \texttt{AEGEAN} where we classified each 2D Gaussian component as a radio source, while \textcite{williams2013} and \textcite{mandal2021} utilise PYthon Blob Detector and Source Finder\footnote{https://github.com/sabourke/pybdsf} (PyBDSF) that groups 2D Gaussians connected via continuous emission into a single source. \textcite{mandal2021} then performed additional steps to deblend physically distinct radio sources that were classified as a single source and consolidate sources with multiple separate radio components (e.g. radio lobes) into a single source. These variations in source classification will result in a discrepancy in source counts between surveys.

The resulting source counts may be adequately described by third-degree polynomials. Fitting the polynomial
\begin{equation}
\begin{split}
\log_{10} \left( S^{2.5} \frac{dN}{dS} \right) = a \log_{10}{(S)}^3 + b \log_{10}{(S)}^2 + c \log_{10}{(S)} + d
\end{split}
\end{equation}

to the GAMA9 and GAMA23 counts, where $S$ is the 215.68\,MHz flux density measured in Jy for the respective fields, results with the constants outlined in Table~\ref{table_poly}.

\begin{table*}[t!]
\begin{threeparttable}
\caption{The resulting constants for a third-degree polynomial fitted to the GAMA9 and GAMA23 source counts.}
\label{table_poly}
\begin{tabular}{lllll}
\toprule
\headrow Field & $a$ & $b$ & $c$ & $d$ \\
\midrule
GAMA9 & $-0.0288 \pm 0.0160$ & $-0.256 \pm 0.040$ & $0.267 \pm 0.032$ & $3.33 \pm 0.02$ \\
\midrule
GAMA23 & $-0.0537 \pm 0.0119$ & $-0.283 \pm 0.030$ & $0.300 \pm 0.024$ & $3.36 \pm 0.02$ \\
\bottomrule
\end{tabular}
\end{threeparttable}
\end{table*}

\subsection{Spectral Properties of the Sample}\label{sample}

\begin{figure}[ht!]
\centering
\includegraphics{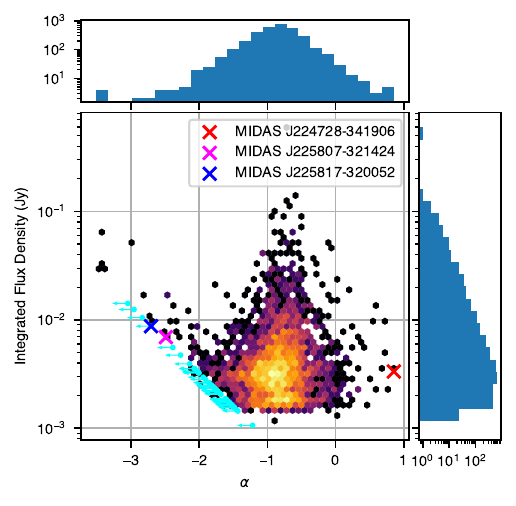}
\caption{The integrated flux density at 215.68 MHz as a function of the spectral indices $\alpha$ between 215.68\,MHz and 887.5\,MHz for sources in the GAMA23 MIDAS catalogue matched to a source within $1'$ of the GAMA23 EMU catalogue \autocite{gurkan2022}. The sources were filtered to unresolved sources along with a selection of sources not detected in EMU as described in the text. The cyan markers are the upper-limits on the spectral index for the selection of sources not detected within EMU. The crosses represent 3 examples of steep spectrum sources, selected to highlight the potential of the catalogue. Here we select one negative spectrum source that is detected in both MIDAS and EMU, one negative spectrum source not detected in EMU and one positive spectrum source.}
\label{fig_spectra}
\end{figure}

\begin{figure*}[hbtp]
\centering

        \begin{subfigure}[b]{\linewidth}
        \centering
        \includegraphics[width=0.392\linewidth]{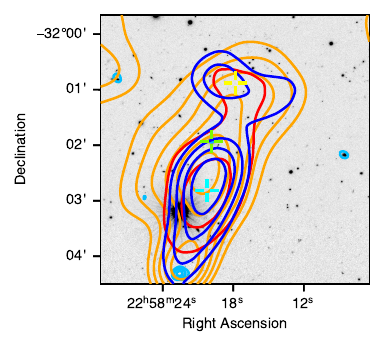} \includegraphics[width=0.465\linewidth]{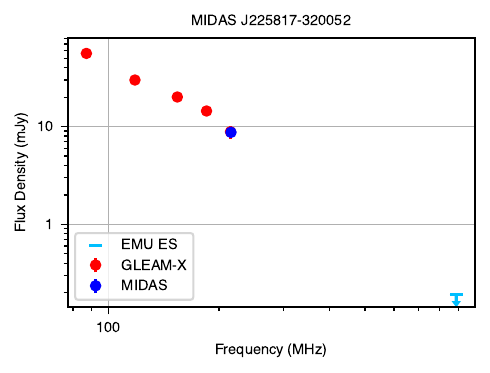}\\
        \caption{MIDAS J225817-320052 with diffuse emission below the detection threshold in EMU. The source is likely one lobe of an asymmetric radio galaxy. The yellow crosshair is the MIDAS lobe not detected in EMU, the green crosshair is the potential host and the cyan crosshair is the the second lobe. A foreground unrelated star is observed in the bottom-left of the optical image.}
        \end{subfigure}

        \begin{subfigure}[b]{\linewidth}
        \centering
        \includegraphics[width=0.392\linewidth]{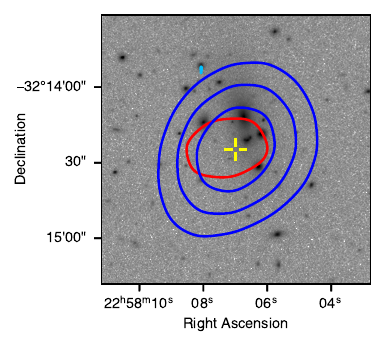}
        \includegraphics[width=0.465\linewidth]{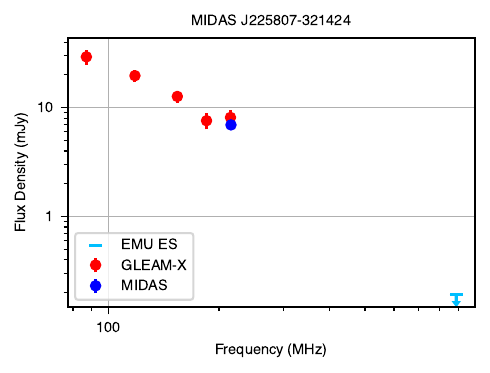}
        \caption{MIDAS J225807-321424, likely diffuse emission from the galaxy cluster GAMAJ225806.74-321411.1 with a point source EMU detection to the top-left just above the $5\sigma$ MIDAS contour.}
        \end{subfigure}

        \begin{subfigure}[b]{\linewidth}
        \centering
        \includegraphics[width=0.392\linewidth]{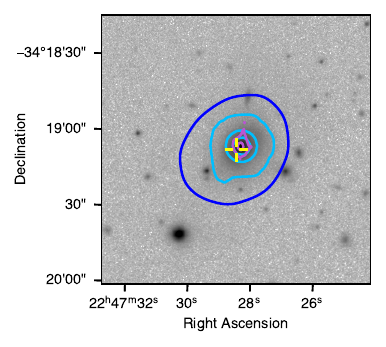} \includegraphics[width=0.465\linewidth]{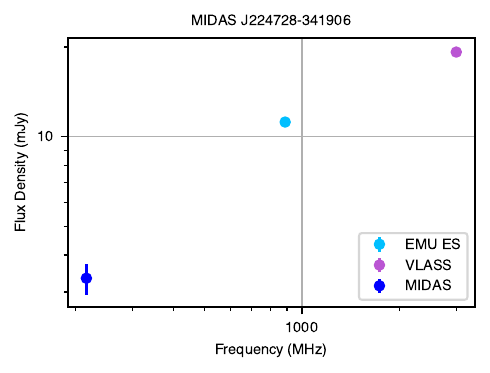}
        \caption{MIDAS J224728-341906, a galaxy with a steep positive spectrum at the low-radio frequencies detected slightly above the $5\sigma$ MIDAS detection threshold. It is very bright at 887.5\,MHz in EMU but is below the detection threshold of GLEAM-X \autocite{ross2024}.}
        \end{subfigure}
        
\caption{Postage stamp radio overlay images of three rare, interesting sources from the GAMA23 field along with their SEDs. The optical background data is $Z$-band imaging from VISTA. The contours are GAMA23 MIDAS 215.68\,MHz (blue), EMU Early Science 887.5\,MHz (sky blue), EMU Early Science 887.5\,MHz convolved to the MIDAS resolution (orange) as described in \S~\ref{rms}, GLEAM-X wideband (170-231\,MHz, red) and VLASS 3\,GHz (purple). The contours start at the reported $5\sigma$ threshold with 50 even increments in log-space to $40\sigma$ for MIDAS J225817-320052 and MIDAS J225807-321424 and three even increments for MIDAS J224728-341906. The yellow crosshairs mark the locations of the sources detected in MIDAS. The SEDs contain the measured integrated flux densities for the sources in each respective survey with the exception of the EMUES values for MIDAS J225818-32005 and MIDAS J225806-32142; as these sources were not detected, we have used their $5\sigma$ detection threshold for the upper limit.}

\label{fig_sources}
\end{figure*}

Sources within the GAMA23 MIDAS catalogue were matched with the 887.5\,MHz deep field source catalogue produced from the ASKAP EMU Early Science project by \textcite{gurkan2022} and the GAMA spectroscopic redshift catalogue \autocite{driver2022}. A maximum separation distance of $1'$ was used for the source matching, resulting in a total of 12~640 sources cross-matched to the EMU catalogue and 5~861 spectroscopic redshifts identified. We detected 846 sources in the GAMA23 MIDAS catalogue within the area imaged by EMU that did not have a corresponding source within $1'$ in the EMU catalogue. Most of these non-detected sources were due to the different source finding methods. \textcite{gurkan2022} used \texttt{PyBSDF} for their source finding and then visually inspected and consolidated radio galaxies with clearly distinct jets and cores into a single source. Our use of \texttt{AEGEAN} attempts to fit 2D Gaussians to the sources, resulting in the core and jets or multiple sections of a complex source being defined as separate sources.

Furthermore, many of the sources not detected in EMU are on the edge of the EMU mosaic in high \ac{RMS} areas, including some sources on the image border, causing them to be cropped. However, there were 60 sources located in the lower RMS areas near the centre of the EMU mosaic that were not detected suggesting spectra with steep negative spectral indices, typical of older emission from remnant radio galaxies or galaxy cluster relics.

From the sources detected in both catalogues, we calculated their spectral indices and identified 35 sources with a spectral index $\alpha^{216}_{888} < -2$ and 5 sources with a spectral index $\alpha^{216}_{888} > 0.7$. These are atypical of standard synchrotron emission from radio galaxies and could point towards interesting physics or objects. In Figure~\ref{fig_spectra} we present the spectral distribution for the point sources calculated between the flux densities in the GAMA23 MIDAS survey at 215.68\,MHz and the EMU survey at 887.5\,MHz along with several of the non-detected sources. For the non-detected sources, the upper limit for the spectral index was calculated from EMU's $5 \sigma$ source detection limit 0.19\,mJy, aligning well with the slope observed for the detected sources in the negative spectral index region. The matched sources were filtered based on the EMU catalogue to those reported as a point source, while non-detected sources were selected from the GAMA23 MIDAS catalogue for a region $8^{\circ}$ by $4^{\circ}$ at the centre of the GAMA23 field to ensure the corresponding area within the EMU image had a low \ac{RMS}. To lower the likelihood of false detections, we filtered out detected sources within $3'$ of sources with an integrated flux density $>0.8$\,Jy  where noise can exceed $5\sigma$. To filter out sources that may be classified differently between the catalogues, the pixel value at each source location was measured, along with the mean of the pixel values for a 20 pixel by 20 pixel region around it. We then selected the sources where these values were below the $5 \sigma$ EMU detection threshold. A brief visual inspection of the 60 sources was performed to verify they were likely genuine.

\begin{table*}[t!]
\begin{threeparttable}
\caption{Details for the three example sources selected from the MIDAS catalogue with extreme spectral indices where 35 sources were identified with a spectral index $\alpha^{216}_{888} < -2$ and 5 sources with a spectral index $\alpha^{216}_{888} > 0.7$. We selected a random source from the negative spectrum population that was detected in both MIDAS and EMU along with a source with one of the steepest negative spectrum that was not detected in EMU. From the positive spectrum population we selected the source with the highest positive spectral index.}
\label{table_spextral_examples}
\begin{tabular}{llll}
\toprule
\headrow & MIDAS J225817-320052 & MIDAS J225807-321424 & MIDAS J224728-341906 \\ \midrule
RA (deg)             & $344.57462$ & $344.52914$ & $341.86905$ \\ \midrule
Dec (deg)            & $-32.01452$ & $-32.24020$ & $-34.31851$ \\ \midrule
$\alpha^{216}_{888}$ & $<-2.7$ & $<-2.5$ & $0.85 \pm 0.07$ \\ \midrule
Int Flux Density (mJy)       & $8.74$  & $6.94$  & $3.34$ \\ \midrule
Peak Flux Density (mJy/beam) & $3.53$  & $6.14$  & $3.19$ \\ \midrule 
Local RMS (mJy/beam) & $0.338$ & $0.344$ & $0.348$ \\ \midrule 
GLEAM-X Source       & GLEAM-X J225817.9-320049 & GLEAM-X J225807.1-321424 & - \\ \midrule 
EMU Source           & - & EMUJ225808.0-321353 (Mismatch) & EMUJ224728.4-341905 \\ \midrule
GAMA Source          & - & GAMAJ225806.74-321411.1 & GAMAJ224728.42-341904.8 \\ \midrule
Redshift             & $0.317$ (Host) & $0.198$ & $0.120$ \\
\bottomrule
\end{tabular}
\end{threeparttable}
\end{table*}

Here we provide a closer inspection for 3 example sources with extreme negative and positive spectral indices detailed in Table~\ref{table_spextral_examples}. From the 35 sources with a spectral index $\alpha^{216}_{888} < -2$, we randomly selected a source that was detected in both surveys (MIDAS J225807-321424) along with one of the brightest sources in MIDAS that was not detected in EMU (MIDAS J225817-320052). From the 5 sources with a spectral index $\alpha^{216}_{888} > 0.7$, we selected the source with the highest spectral index, MIDAS J224728-341906. These sources are detailed in Figure~\ref{fig_sources} using optical data overlaid with the observed radio contours. The optical data is $Z$-band imaging from the Visible and Infrared Survey Telescope for Astronomy (VISTA) Kilo-degree Infrared Galaxy (VIKING) survey \autocite{findlay2012} and the Dark Energy Spectroscopic Instrument Legacy Imaging Surveys \autocite{dey2019}, while the radio contours are from the MIDAS mosaics, GLEAM-X, EMU Early Science and VLASS.

MIDAS J225817-320052 with an integrated flux density of $8.74 \pm 0.34$\,mJy measured in MIDAS was not detected within the EMU mosaic, but is observed in GLEAM-X (GLEAM-X J225817.9-320049) with an integrated flux density of $12.08 \pm 1.74$\,mJy in the wideband mosaic centred at 200\,MHz. Although no source was detected at this source location within the EMU mosaic, diffuse radio emission is observed that is below the EMU detection threshold. The source in MIDAS is likely a radio lobe where a potential host (EMUJ225819.9-320156, GAMAJ225819.84-320154.0) is observed in the EMU mosaic south-southeast to the radio emission with a spectroscopic redshift of $z=0.317$. Additionally, a second brighter source, which is likely the opposing lobe, is observed in MIDAS and GLEAM-X on the opposite side. When convolving EMU to the same beam size as MIDAS, only a large single source, which encompasses the area of the host and lobes, is observed with a peak between the potential host and the southern lobe; this is likely due to the uneven contribution of the radio lobes. The EMU source detected south of the lower lobe appears to be an unrelated background galaxy. Asymmetric double radio sources are unusual. Jet powers are likely similar from either pole of a black hole so the differing morphology is most likely due to environmental affects. For example a denser medium on the southern side could keep a jet collimated longer meaning that we still see a hotspot. Or the southern lobe could be angled away from us so we are seeing it at a younger age than the northern lobe (which would be angled towards us).

MIDAS J225807-321424, randomly selected from the $\alpha^{216}_{888} < -2$ population with an integrated flux density of $6.94 \pm 0.34$\,mJy in MIDAS matched to EMUJ225808.0-321353 with an integrated flux density of $0.205 \pm 0.034$\,mJy resulting in a spectral index of $\alpha^{216}_{888}=-2.5 \pm 0.1$. The MIDAS source position corresponds to the galaxy cluster GAMAJ225806.74-321411.1 \autocite{robotham2011} and has a reported spectroscopic redshift of $z=0.198$ \autocite{driver2022}. This source is also detected in the GLEAM-X wideband mosaic centred at 200\,MHz with an integrated flux density of $8.37 \pm 0.97$\,mJy. The detection within the 887.5\,MHz EMU image occurs north-northeast of the cluster, which appears to be a separate source to the MIDAS detection and therefore is a mismatch resulting in an incorrect spectral index. Within the 215.68\,MHz GAMA23 MIDAS mosaic, the detection occurs near the centre of the cluster, at this location in the convolved EMU mosaic, the radio emission is below the $5\sigma$ detection threshold of 0.190\,mJy. Setting an upper limit for the source within EMU as $5\sigma$, we estimate a spectral index of $<-2.5$

From the sources with positive spectral indices $\alpha^{216}_{888} > 0.7$, we selected MIDAS J224728-341906 which had the highest spectral index in the sample with $\alpha^{216}_{888} = 0.85 \pm 0.07$. It was observed with an integrated flux density of $3.34 \pm 0.35$\,mJy at 215.68\,MHz within the GAMA23 MIDAS mosaic which is too faint to be detected in GLEAM-X. Within EMU at 887.5\,MHz, it was detected with an integrated flux density of $11.2 \pm 0.1$\,mJy and was detected in VLASS at 3\,GHz with an integrated flux density of $19.24 \pm 0.13$\,mJy \autocite{gordon2021}. The source position corresponds to a known galaxy with a spectroscopic redshift of $z = 0.120$ \autocite{driver2022}, previously observed in radio \autocite{wright1996, mauch2013, gordon2021}, infrared \autocite{cutri2013}, optical \autocite{maddox1990} and ultraviolet \autocite{bianchi2011}. The low-frequency MIDAS measurement may be used to help constrain the spectral peak and fit for this source. This source has no reported evidence of variability, so is likely a very young high-frequency peaker. Young radio galaxies peak due to synchrotron self-absorption and that peak moves to lower frequencies as the radio source expands. Future work could model the radio spectra to help determine the mechanism causing the the turn-over.

\section{Conclusion}
 
In this paper, we have presented deep field mosaics produced from the MIDAS survey, with an angular resolution of $\sim60''$, covering the GAMA9 and GAMA23 regions that were observed by the MWA at a central frequency 215.68\,MHz with a bandwidth of 30.72\,MHz. The GAMA9 MIDAS deep field image covers an area ${\sim}1250 \text{\,deg}^2$ centred at RA $+135^{\circ}$, Dec $+0.5^{\circ}$. A total of 1~282 observations were mosaicked, resulting in 42.7\,hours of integration time achieving an RMS of $0.923 \pm 0.003$\,mJy/beam within the GAMA9 region, a depth comparable to RACS and VLASS. The GAMA23 MIDAS deep field image achieved an RMS of $0.331 \pm 0.001$\,mJy/beam within the GAMA23 region from 56.7 hours of integration time (1701 observations) which we believe to be the deepest image produced from the MWA to date. This mosaic covered a ${\sim}850 \text{\,deg}^2$ area centred at RA $+345^{\circ}$, Dec $-32.5^{\circ}$. The low noise achieved at this frequency enables new insight into this field, offering low-frequency radio data to depths comparable to the higher frequency expectant GLASS release and close to the early EMU release.

The MIDAS deep field data complements the wide-field GLEAM-X survey, offering a $2$--$5\times$ improvement in sensitivity for a narrow band centred on 215.68\,MHz in the selected fields of interest. The GAMA23 mosaic approaches the limits of deep field imaging with the MWA, where the asymptotic nature of Figure~\ref{jackknife} suggests the need for longer baselines or additional imaging steps, such as self-calibration or joint deconvolution \autocite{mantovanini2025}, to achieve more significant improvements. The GAMA9 mosaic illustrates the challenges of performing deep surveys with wide-field radio telescopes when there are bright sources within the instrumental response. This highlights the importance in the selection of deep fields for future observations with wide-field radio interferometer telescopes such as the MWA and the SKAO's Low telescope.

The observation processing required to create these mosaics was completed with \ac{DIP}, a pipeline we developed for deep imaging of MWA data based on the data reduction methods of GLEAM-X \autocite{hurley-walker2022}. The pipeline has been optimised to minimise the thermal and sidelobe noise while implementing automation that enables the processing of thousands of observations with minimal user interaction.

From the deep field mosaics, we generated source catalogues adding valuable low-frequency radio data to an existing wealth of shorter wavelength information. From the GAMA9 MIDAS survey, we detected 53~419 sources with a source density of 70.42\,$\text{sources} / \text{deg}^2$ (52.0 beams per source). The catalogue resulting from the GAMA23 MIDAS survey resulted in 80~407 sources with a source density of 140.28\,$\text{sources} / \text{deg}^2$ (32.4 beams per source) which approaches the classical confusion limit. Evaluating the accuracy of the source catalogues, we have demonstrated agreement in the astrometry with NVSS and SUMSS along with RACS-high. Additionally, we demonstrate good agreement in flux density scaling with our reference catalogue GLEAM. A small systematic flux density scale difference between MIDAS and GLEAM-X DRII as a function of SNR is observed. We surmise this is likely due to the $3\sigma$ \texttt{auto-mask} chosen for \texttt{CLEAN}ing in GLEAM-X resulting in a large number of noise peaks being deconvolved and removing flux from true sources.

We have presented the normalised differential source counts calculated from the 215.68\,MHz catalogues. The source counts exhibited good agreement with the 200\,MHz counts by \textcite{franzen2019}, but a deviation below ${\sim} 40$\,mJy with the counts derived from GLEAM-X by \textcite{venville2024}. We attribute this to the lack of a completeness correction applied to their source counts. At ${\sim}150$\,MHz, we observe agreement with \textcite{mandal2021} and the upper estimations of \textcite{mandal2021} from their brightest measurements to ${\sim} 25$\,mJy. At lower flux density limits, we measure higher source counts; this may be due to the difference in resolution and sensitivity of the surveys along with the different source finding methods implemented. 

We matched the GAMA23 MIDAS source catalogue to the 887.5\,MHz GAMA23 source catalogues produced by \textcite{gurkan2022} and identified 35 sources with a spectral index $\alpha^{216}_{888} < -2$ and 5 sources with a spectral index $\alpha^{216}_{888} > 0.7$. We selected a small sample of these sources from which we observed emission from a galaxy cluster, diffuse emissions, and a galaxy not previously detected at low frequencies.

The deep field images and catalogues offer further insight into the low-frequency emission for the regions. A prepublished MIDAS mosaic of the GAMA23 field produced from this work was utilised by \textcite{ighina2024} to determine the 216\,MHz upper limit for a $z{\sim}6.5$ radio-loud quasi stellar object (QSO).

In future work, we will combine our source catalogues with the plethora of data available for the GAMA9 and GAMA23 regions in an attempt to produce accurate \ac{RLF}s. The low-frequency \ac{RLF}s will be important tools for the future study of galaxy evolution.

\begin{acknowledgement}

SP thanks Kat Ross for the assistance provided to understand the GLEAM-X results.

This research is supported by an Australian Government Research Training Program (RTP) Scholarship. N.H.-W. is the recipient of an Australian Research Council Future Fellowship (project number FT190100231).

This scientific work uses data obtained from Inyarrimanha Ilgari Bundara, the CSIRO Murchison Radio-astronomy Observatory. We acknowledge the Wajarri Yamaji People as the Traditional Owners and Native Title Holders of the observatory site. Support for the operation of the MWA is provided by the Australian Government (NCRIS), under a contract to Curtin University administered by Astronomy Australia Limited. We acknowledge the Pawsey Supercomputing Centre which is supported by the Western Australian and Australian Governments.

This research made use of hips2fits,\footnote{https://alasky.cds.unistra.fr/hips-image-services/hips2fits} a service provided by CDS; the NASA/IPAC Extragalactic Database (NED) which is operated by the Jet Propulsion Laboratory, California Institute of Technology, under contract with the National Aeronautics and Space Administration; NASA's Astrophysics Data System and the VizieR catalogue access tool, CDS, Strasbourg, France.

A range of \texttt{python} packages were used for this work including \texttt{pandas} \citep{McKinney_2010, McKinney_2011}; \texttt{wcsaxes}, an open-source plotting library for Python hosted at https://wcsaxes.readthedocs.io/en/latest/; \texttt{Astropy}, a community-developed core Python package for Astronomy \citep{2018AJ....156..123A, 2013A&A...558A..33A}; \texttt{SciPy} \citep{Virtanen_2020}; matplotlib, a Python library for publication quality graphics \citep{Hunter:2007} and \texttt{NumPy} \citep{harris2020array}.

This research made use of \texttt{ds9}, a tool for data visualization supported by the Chandra X-ray Science Center (CXC) and the High Energy Astrophysics Science Archive Center (HEASARC) with support from the JWST Mission office at the Space Telescope Science Institute for 3D visualization and \texttt{TOPCAT}, an interactive graphical viewer and editor for tabular data \citep{2005ASPC..347...29T}. The acknowledgements were compiled with help from the Astronomy Acknowledgement Generator.

This paper includes data that has been provided by AAO Data Central (datacentral.org.au).

Some of the results in this paper have been derived using the HEALPix \citep{gorski2005} package.

\end{acknowledgement}

\paragraph{Data Availability Statement}

The Deep Imaging Pipeline is available from \url{https://github.com/sjpaterson/dip}. The deep field mosaics, RMS maps, PSF maps and catalogues produced in this work for the GAMA9 and GAMA23 regions are available to download from the Data Central PASA Datastore (\url{https://doi.org/10.57891/z19a-2n41}).

\printendnotes

\printbibliography

\appendix
\renewcommand{\thesection}{\Alph{section}}
\renewcommand{\thetable}{\Alph{section}.\arabic{table}}
\renewcommand{\thefigure}{\Alph{section}.\arabic{figure}}
\setcounter{table}{0}
\setcounter{figure}{0}

\onecolumn

\section{Source Masking for Noise Analysis}\label{app_filter}
\begin{figure*}[ht!]
\centering
\textbf{\normalsize{GAMA9}} \\[0.2cm]
\includegraphics[width=0.99\linewidth]{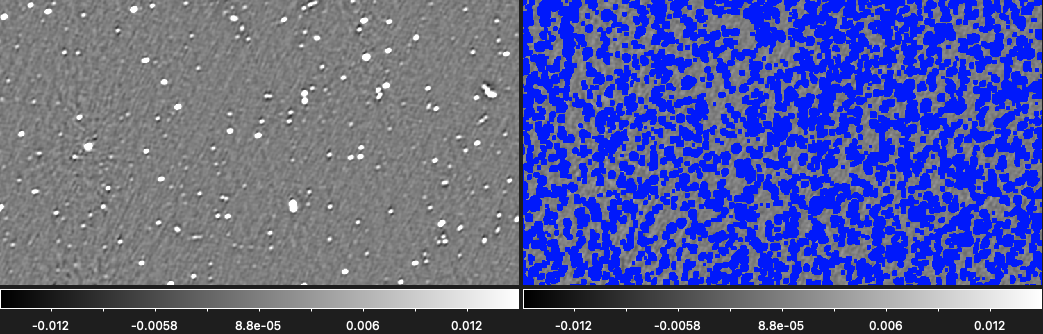} \\
\begin{tabular*}{\linewidth}{@{\extracolsep{\fill}}ccccc}
     & Brightness (Jy/beam) & & Brightness (Jy/beam) & 
\end{tabular*}
\\
\textbf{\normalsize{GAMA23}} \\[0.2cm]
\includegraphics[width=0.99\linewidth]{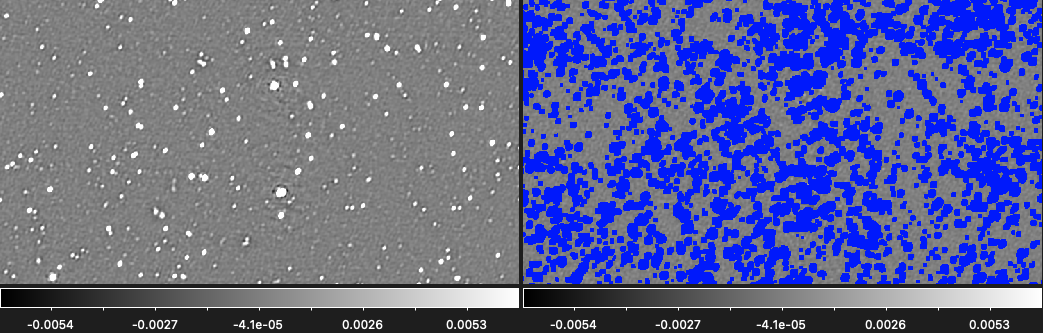} \\
\begin{tabular*}{\linewidth}{@{\extracolsep{\fill}}ccccc}
     & Brightness (Jy/beam) & & Brightness (Jy/beam) & 
\end{tabular*}
\caption{An example of the source masking used for the noise analysis in Section~\ref{rms}. Each cutout covers a 3.8\,deg$^2$ area with the top panels cutout from the MIDAS GAMA9 mosaic and the bottom panels from the MIDAS GAMA23 mosaic. The left-hand panels are cutouts from the mosaic while the right-hand panels are the cutouts with the masking (blue) overlayed.}
\label{app_filter_fig}
\end{figure*}

\newpage

\section{Ionospheric `Blur' Maps}\label{app_blur}
\begin{figure*}[ht!]
\centering
\centering
\includegraphics[width=0.49\linewidth]{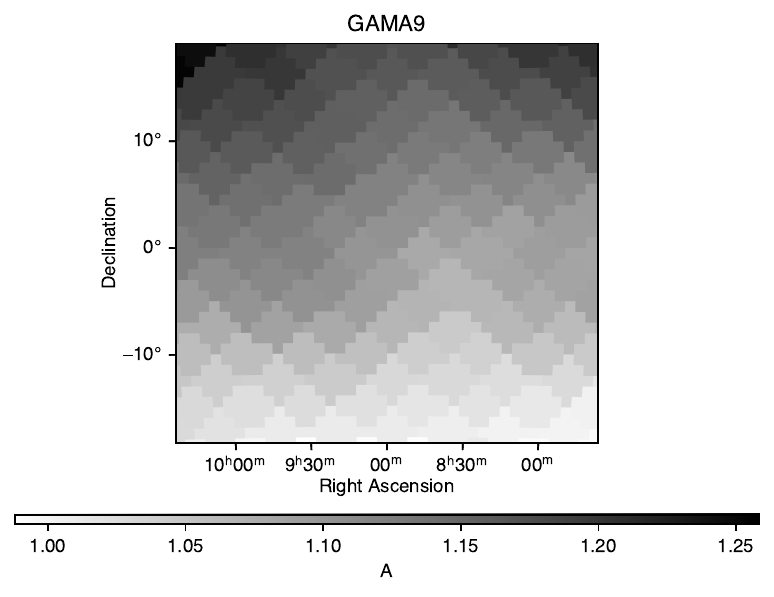}
\includegraphics[width=0.49\linewidth]{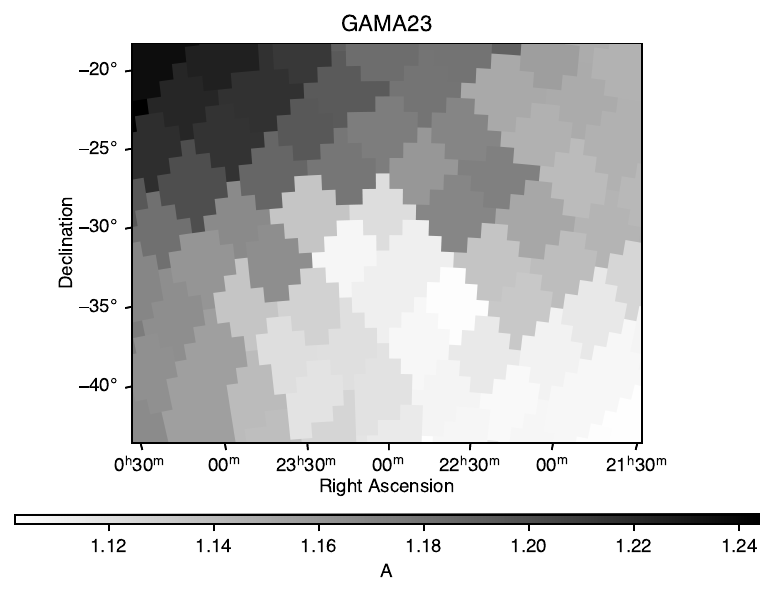}
\caption{The resulting blurring factors (A) calculated in Section \ref{blurring} applied to each MIDAS mosaic to correct the underestimated measured peak flux densities due to ionospheric blurring. Each pixel corresponds to approximately one~square-degree.}
\end{figure*}

\section{Source Catalogue Column Headings}\label{app_colnames}
\setcounter{table}{0}
\begin{table*}[htb!]
\begin{threeparttable}
\caption{The column names, units and description for the 215.68\,MHz MIDAS GAMA source catalogues. The naming follows the coordinate-based naming convention of the International Astronomical Union. The remainder of the columns are the results output by \texttt{AEGEAN}.}
\label{table_columns}
\begin{tabular}{llll}
\toprule
\headrow Number & Name  & Unit & Description \\
\midrule
1 & Name & hh:mm:ss+dd:mm:ss & International Astronomical Union name \\
\midrule
2 & background & Jy/beam & Background flux density level \\
\midrule
3 & local\_rms & Jy/beam & Local RMS \\
\midrule
4 & ra\_str & hh:mm:ss & Right ascension \\
\midrule
5 & dec\_str & hh:mm:ss & Declination \\
\midrule
6 & ra & degrees & Right ascension \\
\midrule
7 & err\_ra & degrees & Uncertainty on the right ascension \\
\midrule
8 & dec & degrees & Declination \\
\midrule
9 & err\_dec & degrees & Uncertaincty on the declination \\
\midrule
10 & peak\_flux & Jy/beam & Peak flux density \\
\midrule
11 & err\_peak\_flux & Jy/beam & Uncertainty for the peak flux density \\
\midrule
12 & int\_flux & Jy & Integrated flux density \\
\midrule
13 & err\_int\_flux & Jy & Uncertainty for the integrated flux density \\
\midrule
14 & a & arcsec & Fitted major axis of the source \\
\midrule
15 & err\_a & arcsec & Uncertainty for the major axis of the source \\
\midrule
16 & b & arcsec & Fitted minor axis of the source \\
\midrule
17 & err\_b & arcsec & Uncertainty for the minor axis of the source \\
\midrule
18 & pa & degrees & Fitted position angle of the source \\
\midrule
19 & err\_pa & degrees & Uncertainty for the position angle of the source \\
\midrule
20 & residual\_mean & Jy/beam & Mean of the residual after source fitting \\
\midrule
21 & residual\_std & Jy/beam & Standard deviation of residual after source fitting \\
\midrule
22 & psf\_a & arcsec & Major axis of the PSF at location of source \\
\midrule
23 & psf\_b & arcsec & Minor axis of the PSF at location of source \\
\midrule
24 & psf\_pa & degrees & Position angle of the PSF at location of source \\
\midrule
25 & reliability & flag & A ``low'' flag for sources within close proximity to a bright source \\
\bottomrule
\end{tabular}
\end{threeparttable}
\end{table*}

\end{document}